\documentclass[useAMS,usenatbib,twocolumn]{mn2e}
\usepackage{xspace}
\usepackage{hyperref}
\usepackage{aas_macros}
\usepackage{url}
\usepackage{amsmath}
\usepackage{color}
\usepackage{amssymb}
\usepackage[capitalise]{cleveref}
\usepackage{graphicx}

\newcommand{\sersic}{S\'{e}rsic\xspace}

\newcommand{\degsq}{\, \mathrm{deg}^2}

\newcommand{\jbca}{{Jodrell Bank Centre for Astrophysics, Department of Physics \& Astronomy, The University of Manchester, Manchester M13 9PL, UK}}

\newcommand{\vla}{VLA\xspace}
\newcommand{\euclid}{\textit{Euclid}\xspace}

\newcommand{\emerlin}{\emph{e}-MERLIN\xspace}
\newcommand{\clean}{\textsc{CLEAN}\xspace}
\newcommand{\pybdsf}{\textsc{PyBDSF}\xspace}

\newcommand{\ih}[1]{\textcolor{black}{#1}}

\title[SuperCLASS -- I. Project overview]{SuperCLASS -- I. The Super CLuster Assisted Shear Survey: Project overview and \ih{ Data Release 1}}
\author[Richard A. Battye {\it et al}]{Richard A. Battye\textsuperscript{\thanks{E-mail: richard.battye@manchester.ac.uk}}$^{1}$, 
Michael L. Brown$^{1}$, 
Caitlin M. Casey$^{2}$, 
Ian Harrison$^{1,3}$, \newauthor
Neal J. Jackson$^{1}$,
Ian Smail$^{4}$, 
Robert A. Watson$^{1}$, 
Christopher A. Hales$^{5,6}$, \newauthor
Sinclaire M. Manning$^{2}$, 
Chao-Ling Hung$^{2}$, 
Christopher J. Riseley$^{7,8,9}$, \newauthor
Filipe B. Abdalla$^{10}$, 
Mark Birkinshaw$^{11}$, 
Constantinos Demetroullas$^{1,12}$, \newauthor
Scott Chapman$^{13}$, 
Robert J. Beswick$^{1}$, 
Tom W.B. Muxlow$^{1}$, \newauthor
Anna Bonaldi$^{1,14}$, 
Stefano Camera$^{1,15,16}$, 
Tom Hillier$^{1}$, 
Scott T. Kay$^{1}$, \newauthor
Aaron Peters$^{1}$, 
David B. Sanders$^{17}$, 
Daniel B. Thomas$^{1}$, 
A.\,P.\ Thomson$^{1}$, \newauthor
Ben Tunbridge$^{1}$, 
Lee Whittaker$^{1,10}$ 
(SuperCLASS Collaboration)
\\
$^{1}$ \jbca\\
$^{2}$ Department of Astronomy, University of Texas at Austin, 2515 Speedway Blvd, Stop C1400, Austin, Texas, U.S.A. \\
$^{3}$ Department of Physics, University of Oxford, Denys Wilkinson Building, Keble Road, Oxford OX1 3RH, UK\\
$^{4}$ Centre for Extragalactic Astronomy, Durham University, South Road, Durham DH1 3LE, U.K.\\
$^{5}$ National Radio Astronomy Observatory, PO Box 0, Socorro, NM 87801, USA\\
$^{6}$ School of Mathematics, Statistics and Physics, Newcastle University, Newcastle upon Tyne NE1 7RU, U.K. \\
$^{7}$ Dipartimento di Fisica e Astronomia, Universit\`a degli Studi di Bologna, via P. Gobetti 93/2, 40129 Bologna, Italy\\
$^{8}$ INAF -- Istituto di Radioastronomia, via P. Gobetti 101, 40129 Bologna, Italy\\
$^{9}$ CSIRO Astronomy and Space Science, PO Box 1130, Bentley, WA 6102, Australia\\
$^{10}$ Department of Physics and Astronomy, University College London, Gower Place, London WC1E 6BT, U.K.\\
$^{11}$ H.H. Willis Physics Laboratory, Tyndall Avenue, Bristol, BS8 1TL, U.K.\\
$^{12}$ Cyprus University of Technology, Archiepiskopou Kyprianou 30, Limassol, 3036, Cyprus\\
$^{13}$ Department of Physics and Atmospheric Science, Dalhousie, Halifax, NS B3H 4R2, Canada\\
$^{14}$ SKA Organization, Lower Withington, Macclesfield, Cheshire SK11 9DL, U.K.\\
$^{15}$ Dipartimento di Fisica, Universit\`a degli Studi di Torino, Via P. Giuria 1, 10125 Torino, Italy\\
$^{16}$ INFN -- Istituto Nazionale di Fisica Nucleare, Sezione di Torino, Via P. Giuria 1, 10125 Torino, Italy\\
$^{17}$ Institute for Astronomy, 2680 Woodlawn Drive, University of Hawaii, Honolulu, HI 96822, U.S.A.}

\begin{document}

\pagerange{\pageref{firstpage}--\pageref{lastpage}} \pubyear{2020}

\maketitle

\label{firstpage}

\begin{abstract}
The SuperCLuster Assisted Shear Survey (SuperCLASS) is a legacy programme using the {\em e}-MERLIN interferometric array. The aim is to observe the sky at L-band (1.4~{\rm GHz}) to a r.m.s. of $7\,\mu{\rm Jy}\,$beam$^{-1}$ over an area of $\sim 1\,{\rm deg}^2$ centred on the Abell 981 supercluster. The main scientific objectives of the project are: (i) to detect the effects of weak lensing in the radio in preparation for similar measurements with the Square Kilometre Array (SKA); (ii) an extinction free census of star formation and AGN activity out to  $z\sim 1$. In this paper we give an overview of the project including the science goals and multi-wavelength coverage before presenting the first data release. We have analysed around 400 hours of {\em e}-MERLIN data allowing us to create a Data Release 1 (DR1) mosaic of $\sim 0.26\,{\rm deg}^2$ to the full depth. These observations have been supplemented with complementary radio observations from the Karl G. Jansky Very Large Array (VLA) and optical/near infra-red observations taken with the Subaru, Canada-France-Hawaii and {\it Spitzer} Telescopes. The main data product is a catalogue of 887 sources detected by the \vla, of which 395 are detected by {\em e}-MERLIN and 197 of these are resolved. We have investigated the size, flux and spectral index properties of these sources finding them compatible with previous studies. Preliminary photometric redshifts, and an assessment of galaxy shapes measured in the radio data, combined with a radio-optical cross-correlation technique probing cosmic shear in a supercluster environment, are presented in companion papers.
\end{abstract}

\section{Introduction}
\label{sec:introduction}

The extragalactic radio source population is a powerful tool to probe a range of astrophysical processes and a number of large scale surveys have been performed covering a wide range of radio frequencies. The two largest (in terms of area) radio source surveys performed at L-band (around $1.4~{\rm GHz})$  are the Faint Images of the Radio Sky at Twenty centimeters (FIRST)~\citep{ 1995ApJ...450..559B} and the NRAO-VLA Sky Survey (NVSS)~\citep{1998AJ....115.1693C} which have each detected $\sim 10^6$ sources and probed the source population to $\sim 1\,{\rm mJy}$ over significant $(>10^4\,{\rm deg}^2)$ areas of the sky. A significant fraction of these sources are AGN/jet-driven sources, but as one goes lower in  flux density it is expected that the radio source population will become dominated by lower luminosity star-forming galaxies (SFGs). This has been confirmed by a number of deeper but much smaller area surveys that have been used to probe the source counts down to flux densities $\sim 20\,\mu{\rm Jy}$, for example, \cite{2008AJ....136.1889O,2017A&A...602A...2S,2010ApJS..188..384S} \citet{1999ApJ...526L..73R,2005MNRAS.358.1159M,2008ApJS..179...95M,2009ApJ...694..235P,2009MNRAS.397..281I,2013MNRAS.436.3759B,2015MNRAS.452.1263P}. Most notably in the present context is a survey of the Hubble Deep Field - North (HDF-N), using the Very Large Array (VLA) and the Multi-Element Radio Linked Interferometer Network (MERLIN)~\citep{2005MNRAS.358.1159M}, which detected a sample of $\sim 100$ radio sources with flux density greater than $40\,\mu{\rm Jy}$ in an area of $\sim 10\,{\rm arcmin}\times 10\,{\rm arcmin}$ with a maximum baseline corresponding to a resolution of $\sim 200\,{\rm mas}$.

The advent of the Square Kilometre Array (SKA) sometime in the next decade will allow one to perform surveys at $\mu{\rm Jy}$ flux densities over significant areas of sky with sub-arcsecond resolution. This will enable a wide-range of science applications as elucidated in \citet{2015aska.confE.174B}. In addition to investigating the nature of the radio sources, and hence the astrophysical processes involved, the increased source density will allow the radio sources to be used as cosmological probes, including weak-lensing surveys \citep{2015aska.confE..23B,2015aska.confE.145B,2015aska.confE..25C,2015aska.confE..18J}. Large L-band surveys with SKA pathfinders -- MIGHTEE with MeerKAT \citep{2016mks..confE...6J} and EMU with ASKAP \citep{2011PASA...28..215N} -- will provide important insights into the source populations for these studies, but cannot measure the small angular scales necessary for detailed morphology studies.

Gravitational lensing is the deflection of electromagnetic radiation by mass concentrations predicted by the theory of General Relativity that can be used to probe the so-called ``dark components" of the Universe - dark matter and dark energy~\citep[e.g.][]{2006IJMPD..15.1753C} - and possible modifications to the theory of gravity~\citep[e.g.][]{2012PhR...513....1C}. Depending on the level of alignment between the source and lensing matter, it can be either strong, whereby multiple images are formed, or weak which leads to the distortion of the shapes of galaxies. This weak lensing effect is coherent across the sky, known as cosmic shear, and is now considered to be one of the primary cosmological tests~\citep{2001PhR...340..291B}. If galaxies are randomly orientated, the cosmic shear two-point correlation function, or equivalently the shear power spectrum, can be inferred by measuring large numbers of galaxy shapes with the main source of statistical uncertainty being due to the dispersion in the intrinsic shapes of the galaxies.

The effects of cosmic shear have been measured in the optical waveband. This started with small areas~\citep{ 1990ApJ...349L...1T,1994MNRAS.271...31M,2000MNRAS.318..625B,2000astro.ph..3338K,2000A&A...358...30V,2000Natur.405..143W} and now the state of the art surveys cover $\sim 100\,{\rm deg}^2$~\citep{2017MNRAS.465.1454H,2018PhRvD..98d3528T,2019PASJ...71...43H}. Future surveys to be performed with the {\it Euclid} satellite~\citep{2011arXiv1110.3193L} and LSST~\citep{2018arXiv180901669T} -- so called stage IV surveys -- should achieve source densities of more than $10\,{\rm arcmin}^{-2}$ over areas of $\sim 10^4\,{\rm deg}^2$. Interestingly, the large-scale surveys performed by the SKA should reach similar levels with sub-arcsecond resolution in the next decade and therefore it is realistic to imagine performing competitive and complementary lensing surveys using the SKA~\citep{2016MNRAS.463.3674H,2016MNRAS.463.3686B,2017MNRAS.464.4747C,2018arXiv181102743S}. 

As already pointed out source densities in the radio waveband are typically much lower than the optical. Nonetheless, cosmic shear has been detected at the $3\sigma$-level in the radio waveband using the FIRST survey with a source density of $\sim 100\,{\rm deg}^{-2}$~\citep{2004ApJ...617..794C}. This pioneering work was followed up in \citet{2010MNRAS.401.2572P} where ellipticities were measured for radio sources detected in the MERLIN/VLA observations of HDF-N. More recently \citet{2016MNRAS.456.3100D,2018MNRAS.473..937D} have found shear correlations between the radio and optical, while \citet{2016MNRAS.463.3339T} and \citet{2019MNRAS.488.5420H} have investigated shear correlations in the COSMOS field, at 1.4~GHz and 3~GHz respectively, with both being limited by the lack of small angular scales observable with the VLA.

The primary science goal of the SuperCLASS project is to detect the effects of weak lensing in the radio band in a survey which is representative of the source population and resolution of that which will ultimately be observed by the SKA. The high-resolution afforded by the combination of two telescopes which have both been recently upgraded -- {\em e}-MERLIN and the Karl G. Jansky Very Large Array (\vla) -- is very similar to that expected for the SKA and has sufficient sensitivity to probe the population of star-forming galaxies expected to dominate SKA surveys, albeit over much smaller areas of the sky. It will allow the development of the tools necessary for shape measurement and a quantitative assessment, for the first time, of the physical properties of the radio sources which can be used for cosmic shear measurements such as the size distribution and the expected r.m.s. ellipticity. Shape measurement techniques in the optical waveband are well developed \citep[e.g.][and references therein]{2013MNRAS.429..661M}, but high resolution radio observations necessarily require an interferometric array. This can be a positive in terms of reducing the impact of the atmosphere, but it also means that the ideas used in optical shape measurement need to be reassessed at the very least and, indeed, the fact that the observations are done in the Fourier domain allows for new possibilities~\citep{2016MNRAS.463.1881R,2018MNRAS.476.2053R,2019MNRAS.482.1096R}.

There are a number of reasons why it is interesting to pursue radio weak lensing:
\begin{itemize}
\item The key instrumental systematic in ground-based optical shear measurements is the stability, both in time and space, of the telescope point spread function (PSF) which has to be deconvolved in order to infer ellipticities. This problem is due to the fact that the observations are affected by atmospheric seeing. In contrast, observations at radio frequencies around L-band  will be diffraction limited and, in principle, the PSF can be calculated directly from the optical properties of the telescope(s) involved.
\smallskip
\item It has been suggested that there is extra information which can be gleaned from the radio observations that can be used to mitigate some of the astrophysical systematics that can hamper cosmic shear measurements. These include the use of polarization and the HI line profile to infer information about the intrinsic orientation of a galaxy and hence mitigate the impact of intrinsic alignments~\citep{2011ApJ...735L..23B,2011MNRAS.410.2057B, 2015MNRAS.451..383W}, as well as the possibility of statistically measuring the redshift distribution of the sources directly from the radio observations~\citep{2017arXiv170408278H}.
\smallskip
\item Whatever observational systematics remain in the radio inferred shape measurements, they will be of very different origin to those present in those from the optical waveband. Therefore, if one were to have radio and optical observations of the same fields from {\it Euclid}, LSST and SKA, as would seem prudent, then it would be possible to cross-correlate the two, removing any uncorrelated systematic effects from the telescopes, such as incorrectly deconvolved PSFs~\citep{2010MNRAS.401.2572P,2016MNRAS.456.3100D,2017MNRAS.464.4747C}.
\end{itemize}
To date, there have been no surveys explicitly designed to measure weak-lensing with radio galaxies. Furthermore, the SKA pathfinders are not suitable for such explorations due to a lack of long baselines accessing small angular scales. This work aims to fill this gap using a specific combination of radio telescopes.

An additional and equally important science goal is to use the fact that radio emission is an unbiased tracer of star-formation to disentangle the effects of star-formation and AGN activity out to $z\sim 1$. This is of particular interest in the context of Dusty Star-Forming Galaxies (DSFGs)~\citep{1997ApJ...490L...5S,2005ApJ...622..772C,2014PhR...541...45C} which have implied star-formation rates (SFRs) of $\sim 100-1000\,{\rm M}_{\odot}\,{\rm yr}^{-1}$ and are believed to contribute significantly to the star formation rate of the Universe at high redshifts. Due to the significant levels of dust obscuration present in these sources, it is not always possible to perform morphological analyses in the optical waveband. Relatively small pilot studies have suggested that such classifications are possible using high resolution radio continuum observations~\citep{2008MNRAS.385..893B,2009MNRAS.399..121C,2009MNRAS.395.1249C,2019ApJ...883..204T}. Fortunately, the survey characteristics necessary to find larger samples, in terms of resolution, depth and area covered, are very similar to those required for weak lensing.

Key to both of these science goals are the unique capabilities of \emerlin in the pre-SKA era. {\em e}-MERLIN is an upgrade of MERLIN that comprises 7 telescopes based in the UK with a maximum baseline length of $217\,{\rm km}$ corresponding to angular scales of $\approx 200\,{\rm mas}$ at L-band~\citep{1986QJRAS..27..413T}. The upgrade has involved replacement of the microwave links which transferred data to the central correlator with fibre optic cables allowing a substantial increase in bandwidth ($512\,{\rm MHz}$ at L-band) and improved receivers; the overall sensitivity improvement when compared to MERLIN at L-band is around a factor of 5. Importantly for the science goals discussed in the next section, {\em e}-MERLIN has the unique ability to resolve star-forming galaxies at these frequencies allowing us to measure their ellipticities and perform a separation of this population from that which is AGN/jet-driven using morphological analysis.

SuperCLASS is part of the {\em e}-MERLIN legacy programme which comprises 12 large projects which have been allocated $\sim50\%$ of the observing time over the first $\sim 5$ years of operation\footnote{See \url{http://www.e-merlin.ac.uk/legacy/projects/} for an overview of these projects}. A number of these programmes have complementary goals to those of SuperCLASS in that they are trying to understand the nature of star/galaxy formation across cosmic time. These ``sibling'' projects include the {\em e}-MERLIN Galaxy Evolution survey (e-MERGE\footnote{\url{http://www.e-merlin.ac.uk/legacy-emerge.html}}, Muxlow et al in prep.) which is performing a very deep $1\,\mu{\rm Jy}$ r.m.s. high resolution survey of the HDF-N following up on \citet{2005MNRAS.358.1159M}, Legacy {\em e}-MERLIN Multi-Band Imaging of Nearby Galaxies (LEMMINGs) whose goal is to make high resolution radio images of a number of nearby galaxies and Astrophysics of Galaxy Transformation and Evolution (AGATE) that will probe the radio source population in the direction of known massive clusters. The combined knowledge gleaned from these projects should lead to a significant advance in understanding the nature of galaxy formation and evolution.

The outline of this paper is as follows: in \cref{sec:observations} we describe the selection of the SuperCLASS field and the rationale behind the multi-wavelength observations which have been taken as part of the project. In \cref{sec:obs} we describe the individual multi-wavelength observations in more detail. \Cref{sec:products} describes the definition of our Data Release 1 (DR1) survey region, the creation of source catalogues for the radio data, and the cross-matching of these sources with catalogues from the optical data. In \cref{sec:science} we describe basic properties of these sources, in particular their sizes, fluxes and \sersic profiles as constrained by the data available to us. \Cref{sec:sumcon} presents our summary and conclusions.

For more detail on the optical redshifts and Spectral Energy Distributions of the radio sources, we refer the reader to \cite{superclass2} (henceforth Paper II) whilst radio and optical shape measurements and weak lensing power spectra constraints are presented in \cite{superclass3} (henceforth Paper III).

\section{SuperCLASS survey}
\label{sec:observations}

The SuperCLuster Assisted Shear Survey (SuperCLASS) is a programme using the {\em e}-MERLIN interferometric array. It was awarded more than 800 hours of observations to perform a deep survey of $\sim 1\,{\rm deg}^2$ as part of the legacy program. This is the first paper from the survey giving an overview of the project and presenting the first data release. It is a companion to two papers presenting the first science results and serves as a reference for the other papers coming from the survey. The first data release comprises around half the observations in terms of time but due to the nature of the observing strategy the resulting mosaic, which we will refer to as ``DR1" throughout this paper, covers $\approx 0.26\,{\rm deg}^2$ to the full depth expected for the survey.  The rest of the observations -- another $\sim 400$ hours -- will bring the survey to full depth over the whole field.
 
\subsection{Survey Design}

The key criteria used in defining the design of the survey were: (i) to have the most circular beam profile possible for both the {\em e}-MERLIN and \vla arrays which will allow the best shape measurement and morphological analysis possible; (ii) to allow efficient observing for telescope arrays located at relatively high latitudes in the northern hemisphere; (iii) have the strongest possible lensing signal over a sufficiently large area to make a meaningful test of shear measurement and to detect the rare counterparts of sub-mm sources. Criteria (i) and (ii) are both met by performing observations at high declinations, while it should be possible to achieve (iii) by observing a region known to have a significant level of large-scale structure, for example, a known super-cluster region~\citep{2016arXiv161204247P}. We note that the choice of a supercluster region also facilitates studies of environmental influences on SF/AGN galaxies, as well as in the background population, as in the STAGES project~\citep{2008MNRAS.385.1431H,2009MNRAS.393.1275G}. It was decided that these considerations outweighed the natural desire to choose one of the commonly observed extragalactic fields which have significant multi-wavelength coverage and hence we have also embarked on a programme of multi-wavelength observations in order to facilitate the science programme. We note that the sensitivity of {\em e}-MERLIN is $\sim 30\%$ higher when observing at high elevation compared to lower declinations necessary to observe more southerly fields.

The choice of depth of the survey and the area covered involves a trade-off between the desire for the survey to be sufficiently deep to detect enough sources to reduce the shot noise in shear measurements and being wide enough to sample the shear correlation mitigating the effects of cosmic/sample variance, and to find rare objects with high star-formation rates. The overall sensitivity is dictated by the survey speed of {\em e}-MERLIN which is a sparsely filled interferometer necessitating significant amounts of integration time. Based on these considerations, and also intending to complement other legacy programs, it was decided to aim for a r.m.s. sensitivity of $\sim 6\,\mu{\rm Jy}$ over an area of $\sim 1\,{\rm deg}^2$ which it was calculated would require 832 hours of {\em e}-MERLIN time including that needed for calibration and using the Lovell Telescope (LT) whose $76\,{\rm m}$ diameter collecting area enables this high level of sensitivity. This should allow high signal-to-noise for detection of sources with flux densities $S>40\,\mu{\rm Jy}$ where we can expect a source density $\sim 1\,{\rm arcmin}^{-2}$. Unfortunately, the inclusion of the LT brings with it an extra complication in that the primary beam of baselines including the LT is much smaller $(\sim 10^{\prime})$ than the primary beams of baselines formed from correlating the other $25\,$m diameter telescopes within {\em e}-MERLIN, which are typically $\sim 30^{\prime}$.

\begin{figure}
\centerline{
\includegraphics[width=0.5\textwidth]{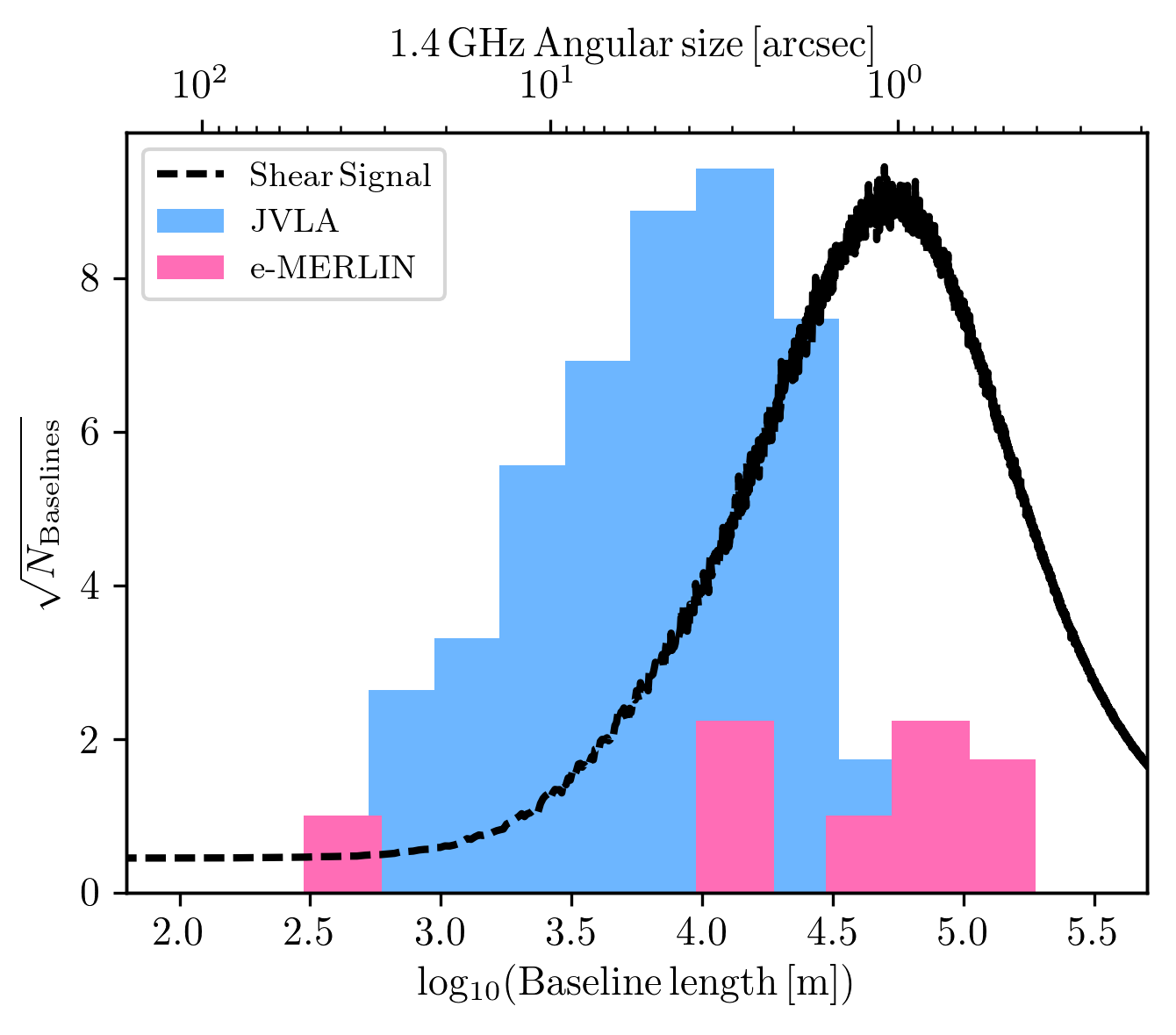}}
\caption{
An illustration of the complementarity of \vla baselines up to $36\,$km and {\em e}-MERLIN baselines up to $217\,$km. Bars represent the square root of the number of baselines of a given length, whilst the dashed line is the shear signal expected on these Fourier scales when observing at 1.4 GHz. The shear signal is constructed as the difference between sky models of T-RECS sources \citep[see][]{2019MNRAS.482....2B} with and without shape changes due to the simulated effect of weak gravitational lensing expected for a typical supercluster of galaxies.}
\label{fig:radiosource}
\end{figure}

As with the survey of the HDF-N presented in \citet{2005MNRAS.358.1159M} it is necessary to complement the {\em e}-MERLIN data with data from the \vla which covers a wider-range of short baselines, but does not have sufficient long baselines to measure ellipticities in relatively small sources. This point is illustrated in Fig.~\ref{fig:radiosource} \ih{where we have compared the lensing signal, computed by differencing a typical source simulated with and without a shear signal added, to the baseline distribution}. Note that the primary beam of the \vla telescopes is similar to the non-LT telescopes of {\em e}-MERLIN and hence it is possible to construct an observing strategy compatible with both arrays.

\subsection{Target field selection}
\begin{figure*}
\centerline{\includegraphics[width=9cm,height=9cm]{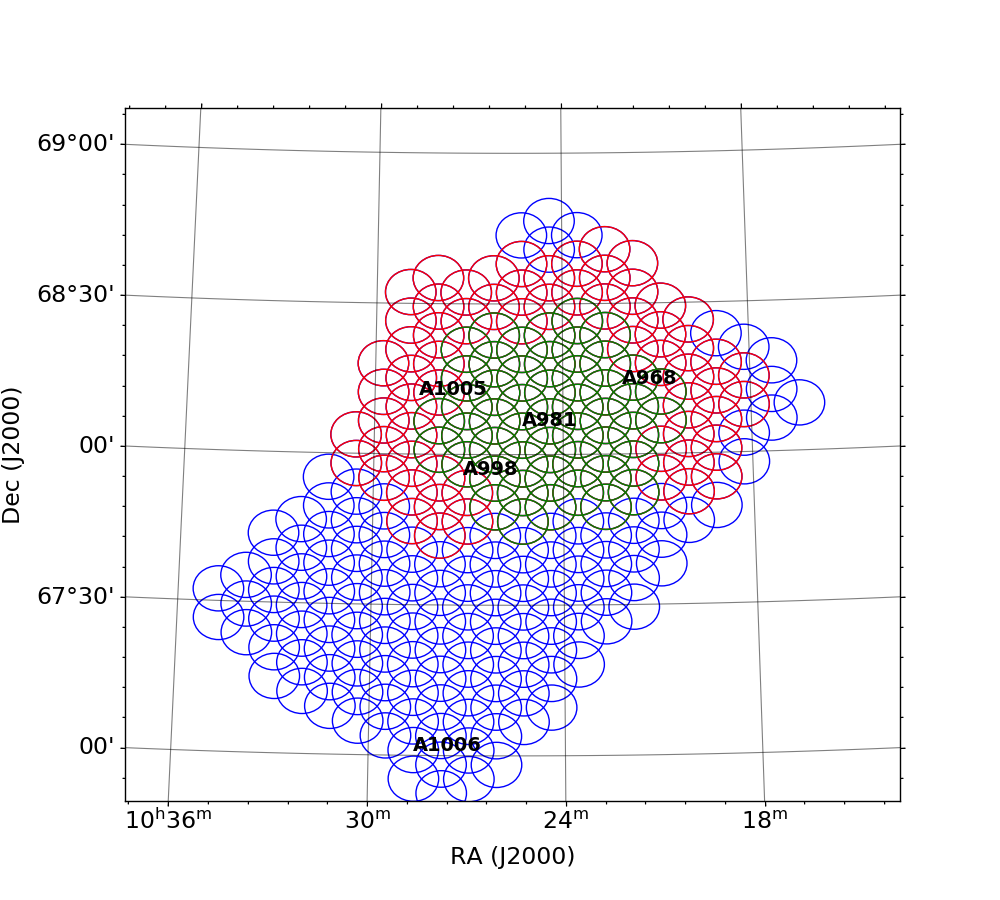}\includegraphics[width=9cm,height=9cm]{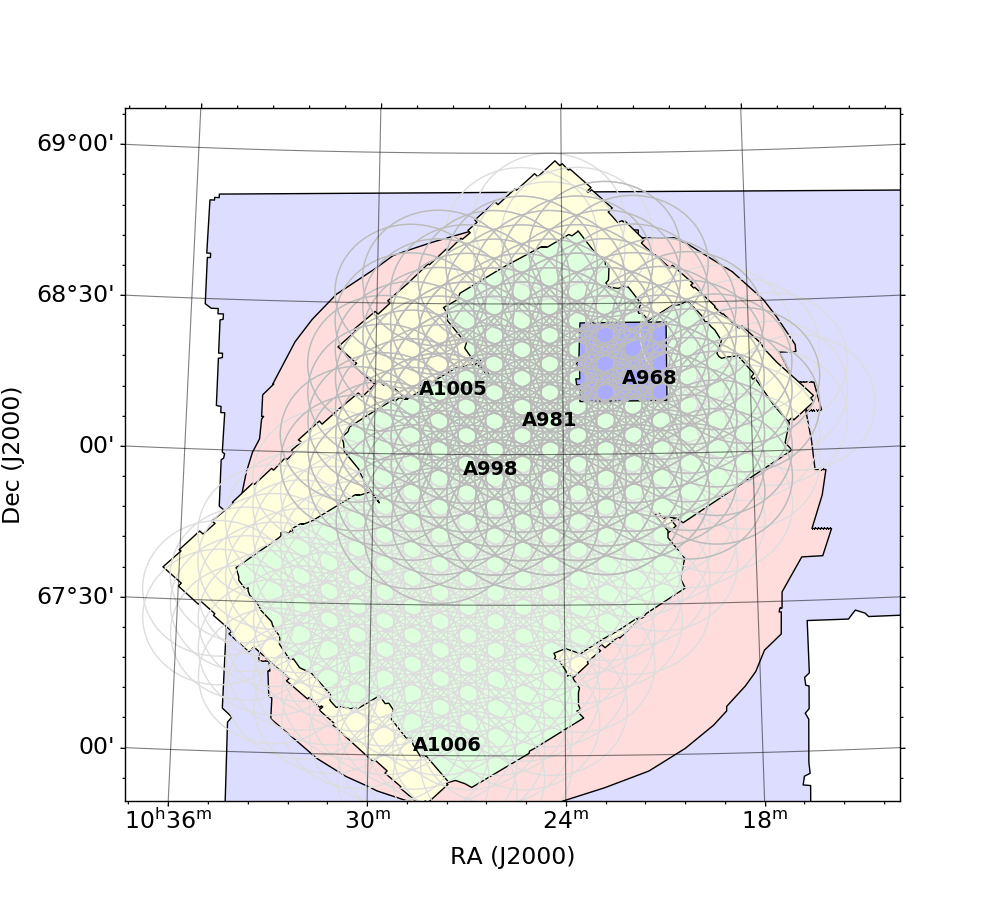}}
\caption{\emph{Left}: the radio pointings for the SuperCLASS field and the positions of the five clusters. The size of the small circles represents the primary beam size of the Lovell Telescope ($\sim 10^{\prime}$) in a hexagonal pattern with a throw of $5.7^{\prime}$. Those in green are the DR1 region discussed in this paper, those in red are the additional SuperCLASS data (which has already been collected), and those in blue are a possible extension. The larger grey circles represent the primary beam of the \vla points (and are indicative of the smaller telescopes in the {\em e}-MERLIN array) which are $\sim 30^{\prime}$ in diameter. \emph{Right}: the coverage of the optical/NIR data with the positions and the radio pointings indicated in grey -- dark grey for those which have been observed with the present time allocation and light grey for the possible extension. The green region is the observed coverage using Suprime-Cam on the Subaru Telescope in bands $BVr^{\prime}i^{\prime}z^{\prime}$ and the yellow region was observed using the same telescope using Hyper Suprime-Cam in the $Y$-band. The purple regions show the coverage using WIRCAM at $K_s$-band using the CFHT and the pink region is that observed at $3.6$ and $4.5\,\mu{\rm m}$ using the IRAC instrument on the {\it Spitzer} Space Telescope. The small purple square represents the region which was not observed by WIRCAM.}
\label{fig:surveyarea}
\end{figure*}
In order to search for possible candidate fields, we performed an all sky search for clusters in the NASA Extragalactic Database (NED) with declinations $>45^{\circ}$, $z>0.15$ and which had previously been studied in order to focus only on actively studied clusters. This list was cross-matched against itself using a matching radius of $0.75\,{\rm deg}$. This process turned up a list of five fields containing more than three clusters. We then excluded regions containing strong sources from NVSS (with flux density $>100\,{\rm mJy}$) and where there is a very strong source $>1\,{\rm Jy}$ within $5\,{\rm deg}$. From those which were left, we selected a region at RA $\sim 10.5\,{\rm h}$ and Dec $\sim 68^{\circ}\,{\rm N}$ containing five clusters (A968, A981, A998, A1005 and A1006, see \cref{tab:clusters}) which have $z\approx 0.2$ -- we will henceforth call this region the SuperCLASS field, since this appears to be the largest of the currently detected clusters in the field. This region has a typical dust extinction of $A_{\rm V}\approx 0.2$ which is within the range that will allow high fidelity optical observations.

The five clusters in the region have been detected by {\it ROSAT} with luminosities in the range $(0.3-1.7)\times 10^{44}\,{\rm erg}\,{\rm sec}^{-1}$ over the 0.1 to $2.4\,{\rm keV}$ energy band as tabulated in Table~\ref{tab:clusters}. Under the assumption of hydrostatic equilibrium, corresponding masses were calculated in \citet{2016arXiv161204247P} which suggest that this region of diameter $\sim 1.5\,{\rm deg}$ contains at least $8\times 10^{14}\,M_{\odot}$ across the five clusters. Using a flat cosmology with Hubble constant $H_0=70\,{\rm km}\,{\rm sec}^{-1}\,{\rm Mpc}^{-1}$ and matter density relative to critical $\Omega_{\rm m}=0.3$, the angular diameter distance to the cluster is $\approx 680\,{\rm Mpc}$ whereas that to typical radio sources with $z\approx 1$ is $\approx 1650\,{\rm Mpc}$. Therefore, the region diameter corresponds to $18\,{\rm Mpc}$ in the supercluster with resolution elements ($\approx 200\,{\rm mas}$) corresponding to $0.6\,{\rm kpc}$ whereas the resolution at typical source redshift is $\approx 1.6\,{\rm kpc}$.

By choosing a supercluster region it will be possible to study the environmental influences on SF/AGN galaxies as well as in the background population as in the STAGES project~\citep{2008MNRAS.385.1431H,2009MNRAS.393.1275G}.

\begin{table*}
\caption{Physical characteristics of the clusters in the SuperCLASS field. The masses are estimated on the basis of hydrostatic mass and a scaling relation between mass and the observed X-ray luminosity $L_{\rm X}$ and hence are only indicative as mass measurements. The total mass in the supercluster is estimated to be $>8\times 10^{14}M_{\odot}$. For details we refer the reader to  \citet{2016arXiv161204247P} and references therein. Note that Abell 1006 is not in the region of the supercluster field that will be observed as part of the presently awarded observation time.}
\label{tab:clusters}
\begin{center}
\begin{tabular}{cccccc}
\hline
Name & RA(B1950) & Dec(B1950) & $z$ & $L_{\rm X}/10^{44}\,{\rm erg}\,{\rm sec}^{-1}$ & $M_{500}/10^{14}M_{\odot}$ \\
\hline
Abell 968 & $10^{\rm h}17^{\rm m}44.1^{\rm s}$ & $+68^{\circ}36^{\prime}34^{\prime\prime}$ & 0.195 & 0.4 & $1.2\pm 0.3$ \\
Abell 981 & $10^{\rm h}20^{\rm m}36.0^{\rm s}$ & $+68^{\circ}20^{\prime}06^{\prime\prime}$ & 0.201 & 1.7 & $2.7\pm 0.7$ \\
Abell 998 & $10^{\rm h}22^{\rm m}47.8^{\rm s}$ & $+68^{\circ}11^{\prime}13^{\prime\prime}$ & 0.203 & 0.4 & $1.2\pm 0.3$ \\
Abell 1005 & $10^{\rm h}23^{\rm m}40.0^{\rm s}$ & $+68^{\circ}27^{\prime}18^{\prime\prime}$ & 0.200 & 0.3 & $1.0\pm 0.2$ \\
Abell 1006 & $10^{\rm h}24^{\rm m}10.7^{\rm s}$ & $+68^{\circ}17^{\prime}44^{\prime\prime}$ & 0.204 & 1.3 & $2.4\pm 0.6$ \\
\hline
\end{tabular}
\end{center}
\end{table*}

We have already pointed out that the inclusion of the LT complicates our observations since its primary beam is about a factor of 3 smaller than the other telescopes in the {\em e}-MERLIN array and those which are part of the \vla. In order to cover the SuperCLASS field with approximately uniform noise coverage, it is necessary to use a hexagonally orientated mosaicing strategy whose pointing centres are separated by $5.7^{\prime}$. The radio pointings observed as part of the project are illustrated in the left hand panel of Fig.~\ref{fig:surveyarea} where the coloured circles are indicative of the primary beam of the LT and the grey circles are the equivalent for the smaller {\em e}-MERLIN telescopes and those in the \vla. The observations of the DR1 region presented in this paper are coloured in green. The red circles indicate the additional SuperCLASS pointings for which the observations are now complete and the data are currently being analysed. The full field covers $\approx 1\,{\rm deg}^2$, made up of a total of 112 pointings. This region includes 4 of the clusters (A968, A981, A998 and A1005) which comprise most of the known mass in the region. We have also included in Fig.~\ref{fig:surveyarea} a proposed extension to the south which would double the area and include the other cluster (A1006). It may be possible to observe this region in a fraction of the time using the phased array feed that will soon be installed on the LT.

\begin{figure}
\centerline{\includegraphics[width=8cm,height=6cm]{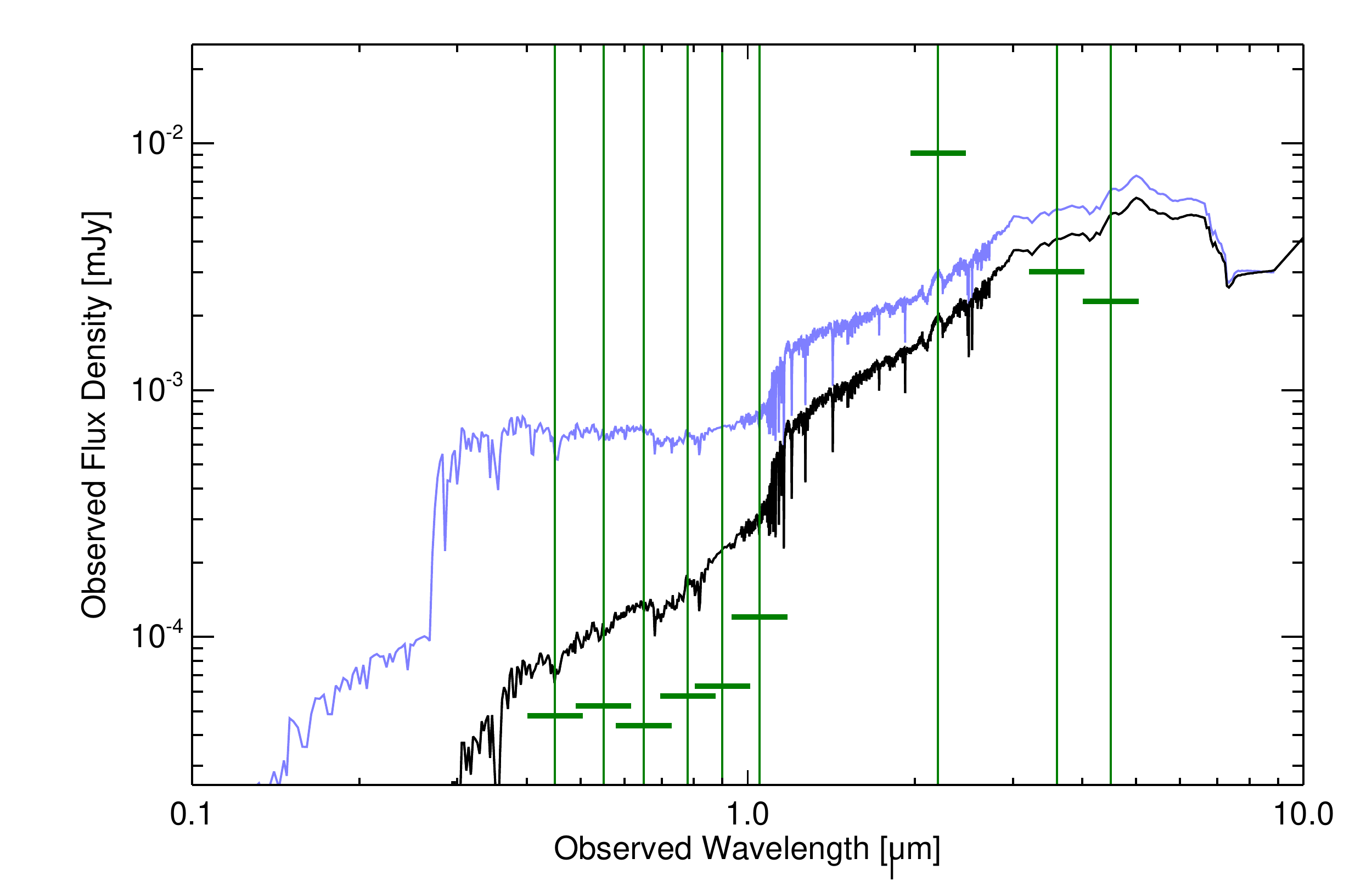}}
\caption{A typical spectral energy distribution for a moderate-age dusty star-forming galaxy at $z=1.5$, without inclusion of nebular line emission. In the blue is the SED that would be observed under the unrealistic assumption that it was not affected by dust obscuration and in black is the same spectrum as in blue, but processed using standard dust attenuation models~\citep{2001PASP..113.1449C} combined with synthetic modelling of stellar populations~\citep{2003MNRAS.344.1000B}. The vertical green lines indicate the positions on the SED of the bands which we have observed. These are $BVr^{\prime}i^{\prime}z^{\prime}Y$-bands in the range $\sim 0.4-1.0\mu{\rm m}$ from Subaru, $K_s$ $\sim 2\,\mu{\rm m}$  from CFHT and $3.6/4.5\,\mu{\rm m}$ from {\it Spitzer}. The horizontal lines denote the 90\% completion depths of the observations presented as part of this first data release which are summarised in Table~\ref{tab:depth}.}
\label{fig:base_sed}
\end{figure}

\section{Observations}
\label{sec:obs}
The field we have chosen to observe with {\em e}-MERLIN and the \vla is not one of the commonly observed extragalactic fields and therefore, in addition to the radio observations, we have embarked on an extensive programme of optical/NIR observations. These are necessary to obtain photometric redshifts for the detected radio sources. Estimated redshifts are absolutely essential for both of the main science goals of the project: for the lensing observations they are necessary to separate the background and foreground galaxies and to estimate the expected signal from theory, while in the context of galaxy formation/evolution applications they are essential for estimating the size and luminosity of observed sources.  

The typical $r^\prime$-band magnitude for the sources we expect to detect using radio observations is $r^\prime \approx 22-25$~\citep{2004ApJ...603L..69C}. The spectral energy distribution (SED) of a typical DSFG at $z=1.5$ is presented in Fig.~\ref{fig:base_sed} along with the wavebands we have observed so far and which are discussed in Section~\ref{sec:obs}. These wavebands are the $BVr^\prime i^\prime z^\prime$-bands observed using the Suprime-Cam on the Subaru Telescope with a goal $AB$ magnitude limit of 25, $Y$-band using Hyper Suprime-Cam again on Subaru to $AB<25$, $K_s$ band on the Canada-France-Hawaii Telescope using WIRCAM with a goal of $AB<22$, and observations at wavelengths $3.5$ and $4.6\mu{\mathrm m}$ using the IRAC instrument on the {\it Spitzer} Space Telescope. These observing bands straddle the $4000\AA$ break in the spectrum over a wide range of redshifts, including that presented in Fig.~\ref{fig:base_sed}, suggesting that it should be possible to obtain a estimate of the photometric redshift distribution at least in a statistical sense. The footprints of the optical/NIR observations presented in this paper are shown in the right hand panel of Fig.~\ref{fig:surveyarea}.

As well as contributing to constraining photometric redshifts, the observations made by {\it Spitzer} can facilitate other science. Firstly, they will allow estimates of the stellar masses for galaxies at high redshift that should allow the separation of contributions to radio emission due to starburst activity. Moreover, since AGN radio emission emanates from areas $\ll1\,{\rm kpc}$, while radio emission from star-forming regions will emanate from the entire galaxy ($\sim 3-10\,{\rm kpc}$), it will be possible to use the mid-infrared power-law between $3.6$ and $4.5\mu{\rm m}$, combined with the concentration of the radio emission to separate star-formation from AGN activity.

\subsection{{\em e}-MERLIN observations}

The {\em e}-MERLIN observations were performed over the frequency range $1.204-1.717\,{\rm GHz}$. The bandwidth was divided into 8 Intermediate Frequency (IF) bands, each of 512 channels of width $125\,{\rm kHz}$. This channel width gives a field of view with acceptably low bandwidth smearing of about $30^{\prime}$, sufficient to map the whole primary beam of the smallest ($25\,{\rm m}$ diameter) telescopes in the {\em e}-MERLIN array. The integration time used was $1\,{\rm sec}$ throughout; the choice was again made to maintain acceptably low smearing in the image across $30\,$arcmin. The data were not averaged further in either time or frequency during the analysis. Observations began on December 29th 2014 and continued with further observing seasons in 2015 and 2016.

Observations were performed in units of 15 hours. Each observation consisted of eight passes over seven grid points in a hexagonal configuration. The resulting 56 scans lasted 12.5 minutes and were interspersed with a 2.5 minute observation of a phase calibrator. The phase calibrator used was J1034+6832, a point source of flux density $130\,{\rm mJy}$ at a distance of about  $1\,{\rm deg}$ selected from the JVAS survey~\citep{1992MNRAS.254..655P}. Each observation was repeated four times, with the starting hour angle of each repetition chosen to ensure as complete $uv$ coverage as possible in the combined field. Each pointing was therefore observed for 400 minutes in total. Combination of pointings into mosaics, with the $5.7^{\prime}$ pointing separation, is expected theoretically to increase the signal-to-noise by a factor of 2, and this factor was achieved in practice. The resulting noise levels in the mosaiced images are typically $7\,\mu{\rm Jy}$ which is close to the goal sensitivity for the survey.

In addition to the phase calibrator observations every 15 minutes, the bright flux calibration source 3C286 was observed once per 15~hour observation. This was done to ensure an accurate overall flux density scale \citep{2013ApJS..204...19P}. OQ208, a bright point source, was also observed for the purpose of bandpass calibration. Finally, 3C84, an extremely bright point source (about $20\,{\rm Jy}$ at L-band) was observed as an instrumental polarization calibrator, since it has almost zero intrinsic polarization. We defer discussion of the polarization results to a later paper.

Data were reduced in the NRAO {\tt AIPS} package using a modified version of the pipeline described by \citet{2014ascl.soft07017A}. The data were cleaned of Radio Frequency Interference (RFI) using a version of the {\tt AOFlagger} program \citep{2010MNRAS.405..155O,2010rfim.workE..36O} whereby binned samples which were above set thresholds in amplitude and noise were flagged. About 15\% of the data were typically lost in this way, with the lowest-frequency IF being most affected. Data were then fringe-fitted to remove instrumental delays; although these were usually small, in some epochs delays of $100-200\,{\rm nsec}$ were seen, which underwent sudden changes during the observations due to reconfigurations within the correlator. Bandpass corrections were then performed, together with an initial phase calibration using the phase calibrator source J1034+6832, and, if necessary, an extra round of self-calibration. Phase corrections were applied to the data and an overall visibility clip at $8\,{\rm Jy}$ was performed on the target field to exclude aberrant data or data still affected by residual RFI. 

Strong sources were imaged and removed from the data by a ``peeling'' process  \citep{2004SPIE.5489..817N}. In the case of moderately strong sources $< 1\,{\rm mJy}$, the data were phase shifted to the source position, the source was then imaged and the {\tt CLEAN} components were subtracted from the data. In the case of very strong sources, the source was imaged without phase shifting and the data self-calibrated using the very strong source. This source was then subtracted, and the self-calibration solutions unapplied from the data. About 10 sources needed to be rigorously removed from nearby pointings in order to eliminate their sidelobes from the images of other sources. Data were then exported to {\tt FITS} files and read into {\tt CASA} measurement sets and the single pointing imaging was performed using the {\tt WSCLEAN} algorithm~\citep{2014MNRAS.444..606O}. Cotton-Schwab cleaning with {\tt mgain} 0.9 was used with natural weighting for optimal noise and diffuse source sensitivity. Cleaning was stopped when clean components of $45\,\mu{\rm Jy}$ were reached. The full mosaic image was compiled by using the {\tt AIPS FLAT} task to combine the individual pointings using the primary beam correction parameters from the original {\em e}-MERLIN pipeline.

\subsection{\vla observations}

We used the \vla in its A configuration to observe all 112 pointings over the frequency range $1-2\,{\rm GHz}$ (L-band). The observations were performed in August 2015 under project code 15A-053. The total observing time was 24 hours, divided into 6$\times$4~hour sessions. These were assigned staggered local sidereal time start times to maximize coverage in the $uv$ plane. All 112 fields were observed in each session. The correlator was configured to deliver $250\,{\rm kHz}$ spectral channels (a total of 4,000) in full polarization with $1\,{\rm sec}$ time sampling. We observed 3C286 for the purpose of bandpass, flux density, and position angle calibrations, J0954+7435 for amplitude and phase calibrations, and the unpolarized source J1400+6210 for instrumental leakage calibration. J1400+6210 is less than 0.05\% polarized at L-band.

The data were calibrated using version 4.7.0-1 of the {\tt CASA} package~\citep{2007ASPC..376..127M}. Hanning smoothing was performed. RFI was identified and flagged using a combination of manual inspection for 3C286 and automated processing for all other targets using {\tt pieflag}~\citep{2014ascl.soft08014H}. The data were flagged to ensure a consistent cross hand phase frame using the task {\tt antintflag}~\citep{hales:antintflag}. GPS-derived total electron content (TEC) data were obtained from the International GNSS Service and used in combination with the International Geomagnetic Reference Field model of the Earth's magnetic field to account for atmospheric Faraday rotation when calculating and applying calibration solutions. Line of sight TEC values throughout the observations ranged between 15 to 35 TECU, translating to rotation measures $<4\,{\rm rad}\,{\rm m}^{-2}$. The flux density scale was referenced to the most recent 2012 value for 3C286 from \citet{2013ApJS..204...19P}. The position angle was assumed to be $-33^{\circ}$ following \citet{2013ApJS..206...16P}. Gain solutions were interpolated using {\tt interpgain} \citep{hales:interpgain} to avoid excessive flagging in {\tt applycal} when individual antennas were periodically unavailable due to RFI flagging. Overall, 38\% of the target field data were flagged. The quality of the polarization calibration was assessed by imaging the calibrated data for 3C286 per frequency channel. The recovered spectra for position angle and fractional polarization from each observing session were found to be consistent with \citet{2013ApJS..206...16P}, modulo a  0.3\% increase in fractional polarization due to the higher spatial resolution of our data compared to their D configuration data.

The data were imaged using {\tt CASA} version 4.7.2. Each of the 112 pointings were imaged individually over a $1.4\,\degsq$ field-of-view with $0.2^{\prime\prime}$ square pixels and common resolution $1.9^{\prime\prime} \times 1.5^{\prime\prime}$ with position angle $80^{\circ}$ east of north. Imaging was performed using a dual Taylor term expansion about a reference frequency of 1.5 GHz. Briggs weighting was employed with robust parameter unity to achieve near natural limited sensitivity with an improved, and closer to Gaussian, beam. Widefield corrections were included to account for the non-coplanar nature of the array. A single round of phase self-calibration was applied to each pointing using a manual selection of compact sources. In 49 pointings containing strong extended emission, this required cleaning and subtracting all other sources in the field prior to performing self-calibration using the intended compact sources. This process yielded on-axis r.m.s. noise levels less than $20\,\mu{\rm Jy}\,{\rm beam}^{-1}$ in all 112 individual pointings. The images were corrected for primary beam attenuation, truncated at the half-power radius, and then linearly mosaiced in slant orthographic (SIN) projection using version 1.5 of the MIRIAD package \citep{1995ASPC...77..433S}. The r.m.s. noise is typically less than $7\,\mu{\rm Jy}\,{\rm beam}^{-1}$ across the final mosaic, except for small regions around some bright sources (tens of mJy). Polarization imaging and analysis will be described in a later paper.

\subsection{Subaru Optical Imaging}
\label{sec:subaru}
Observations of the SuperCLASS field were taken using the Suprime-Cam (SC)  and Hyper Suprime-Cam  (HSC) instruments on the Subaru $8.4\,{\rm m}$ Telescope. SC observations covering the $BVr^{\prime}i^{\prime}z^{\prime}$ filters comprising six fields, each with a field-of-view of $34^{\prime}\times 27^{\prime}$, were taken over the nights of 5th and 6th February 2013. A significant amount of time on the second night was lost due to poor weather resulting in shorter exposure times in the $z^\prime$-band and indeed no coverage for the northern part of the field which has significant overlap with the DR1 region. The total exposure times ranged from 6 minutes ($z^\prime$-band) to 33 minutes ($i^\prime$-band) and the average seeing range from $1.04^{\prime\prime}$ to $1.38^{\prime\prime}$. The HSC data with the $Y$-band filter were taken on 27th March 2015 in two fields, north and south, whose exposure times were 106 minutes and 96 minutes respectively and had a seeing of $1.2^{\prime\prime}$.

The individual exposures have been used for optical shape analysis using the best seeing possible as presented in~\citet{superclass3} (Paper III) but in order to derive a photometric catalogue for photometric redshift estimation we PSF match all the images removing differences in the seeing and exposure time. In order to do this we use the SC image with the worst seeing ($1.38^{\prime\prime}$) as our reference image, create a kernel describing the difference in seeing, and then degrade the PSF of the image for each filter to match the reference image. This convolution smooths noise and signal, so that only contrast and not signal-to-noise is lost.

Flux calibration is performed by comparison to standard stars within the fields. Since the field is at high galactic latitude the Guide Star Catalog (GSC) is incomplete and therefore we have used a sample of stars which are unsaturated in both the SuperCLASS observations and those from the Data Release 1 (DR1) of the Panoramic Survey Telescope and Rapid Response System (PanSTARRS) $3\pi$ survey~\citep{2016arXiv161205560C} which have similar filter bands. First, stars are separated by comparing the maximum surface brightness with the total magnitude which will have a constant ratio for point sources (stars). The PanSTARRS magnitudes are then ``colour-corrected'' using stellar templates from the Stellar Flux Library \citep{1998PASP..110..863P} which are convolved with the Subaru filters. Comparison of the magnitudes from SuperCLASS and PanSTARRS then allows a simple zero-point magnitude calibration of the images. 

This process yields a photometric catalogue with $\sim 6\times 10^{5}$ sources in the $BVr^\prime i^\prime$-bands, around $2.9\times 10^{5}$ in the $z^\prime$-band - which was affected by the bad weather on the 2nd night of observations, and $1.1\times 10^5$ in the $Y$-band once multiple detections from overlapping fields are removed. The limiting $AB$ magnitudes for $90\%$ and total numbers of detected sources across the 6 Subaru bands are presented in Table~\ref{tab:depth}. We see that they are close to the goal of $AB=25$.

For the science analyses done as part of the DR1 we have performed cuts on the full photometric sample in order to reduce the number of artefacts. In Paper III we define a subsample -- the optical lensing sample -- containing 106934 sources corresponding to a source density of $\sim 19\,{\rm arcmin}^{-2}$ in the $i$-band used for the optical lensing analysis presented in that paper.

\begin{table*}
\caption{Summary of the optical/NIR observations discussed in sections~\ref{sec:subaru},~\ref{sec:cfht} and \ref{sec:spitzer}. We have included an estimate of the centre wavelength of the filter, the $AB$ magnitude for $90\%$ completion and the number of sources detected. There is relative uniform coverage across $BVr^\prime i^\prime$-bands close to the goal $AB$ magnitude limit of 25. $z^\prime$-band observations were curtailed by bad weather. Note that there are significantly fewer sources detected in the $Y$-, $K_s$-, $N$- and $M$-bands.}
\label{tab:depth}
\begin{center}
\begin{tabular}{ccccc}
\hline
Telescope/Instrument & Band & ${\rm Wavelength}/\mu{\rm m}$ & $90\%$ completion limit & Number of sources \\
\hline
Subaru/SC & $B$ & 0.45& 24.8 & 376173 \\ 
Subaru/SC & $V$ & 0.55  & 24.6 & 376059 \\ 
Subaru/SC & $r^\prime$ & 0.65 & 24.6 & 375978 \\ 
Subaru/SC & $i^\prime$ & 0.80  & 24.5 & 375728 \\ 
Subaru/SC & $z^\prime$ & 0.90 & 24.5 & 193442 \\ 
Subaru/HSC & $Y$ & 1.0 & 23.8 & 117318 \\ 
CFHT/WIRCAM & $K_s$ & 2.2 & 19.0 & 47819 \\
{\it Spitzer} & $N [3.6]$ & 3.6 & 20.0 & 51364 \\ 
{\it Spitzer} & $M [4.5]$ & 4.5 & 20.1 & 54137 \\ 
\hline
\end{tabular}
\end{center}
\end{table*}

\subsection{Canada-France-Hawaii Telescope observations}
\label{sec:cfht}
 Near-IR $K$-band observations of the SuperCLASS field were obtained with WIRCAM on the Canada-France-Hawaii Telescope (PI Chapman). Given the field of view of the instrument (20$\arcmin\times$20$\arcmin$), imaging was mosaiced across the field in a variety of conditions spanning multiple nights in queue mode. Images reach moderate depth (AB\,$<$\,19), but unfortunately a key section of the field (encompassing the DR1 coverage area) lacks coverage. Because of this gap and relatively shallow and non-uniform depth, K-band is not included in the photometric redshift analysis.
\subsection{{\it Spitzer} IRAC observations}
\label{sec:spitzer}
Time was awarded as part of {\it Spitzer's} Cycle 12 in program \#12074 (PI Casey) to survey the SuperCLASS field. The field was observed for 12.7 hours with two pointings using the IRAC instrument in $M$-band with central wavelength of $3.6\,\mu{\rm m}$ and $N$-band at $4.5\,\mu{\rm m}$. Flux calibration was completed using astronomical standard stars in {\it Spitzer}'s continuous viewing zone and were monitored throughout observations.  We found the ``Post-BCD'' (or post-basic calibrated data) software data products sufficient for our scientific use. Sources were extracted in the {\it Spitzer} bands using \textsc{SExtractor} \citep{1996A&AS..117..393B}, accounting for the large point response function (PRF) of the warm mission, 1.78$''$ for the $M$-band and 1.80$''$ for the $N$-band.  No prior positions were used for {\it Spitzer} source extraction and detections were treated as independent from the source identified in the optical bands.  We obtained $90\%$ completion magnitudes of 19.2 and 19.7 for the $M$- and $N$-bands respectively, finding 3.7$\times10^{4}$ and 3.8$\times10^{4}$ sources in the two filters respectively covering an area of nearly 2\,deg$^2$.  Sources are matched to the optical catalogue with a nearest neighbour approach up to a maximum separation of 0.1$''$.

\subsection{Ancillary multi-wavelength observations of the SuperCLASS field}

We have also observed the SuperCLASS field in a number of additional wavebands whose results will be discussed in other papers. 
\begin{itemize}
\item Observations of the field using the Giant Metrewave Radio Telescope (GMRT) at $325\,{\rm MHz}$ were reported in \citet{2016MNRAS.462..917R}. These achieved a r.m.s. noise level of $34\,\mu{\rm Jy}\,{\rm beam}^{-1}$ over an area of $\approx 6.5\,{\rm deg}^2$ and detected a total of 3257 sources with flux densities between $183\,\mu{\rm Jy}$ and $1.5\,{\rm Jy}$. There are 454 sources in the SCG325 catalogue that lie within the DR1 field. We will consider matches between these sources and those detected using the {\em e}-MERLIN/\vla observations in the DR1 region in Section\,\ref{sec:science}.
\item The SuperCLASS field was observed with the Low-Frequency Array \citep[LOFAR;][]{2013A&A...556A...2V} High-Band Antennas (HBA) during Cycle 0 on 3rd April 2013, using the Dutch array (core and remote stations only, no international stations). The field was covered with a single pointing centered on Abell 998, for a total integration time of approximately eight hours. The frequency range covered is $115-162\,{\rm  MHz}$. We expect to achieve approximately $\approx 50-100\,\mu{\rm Jy}\,{\rm beam}^{-1}$ sensitivity with LOFAR \citep[based on similar observations by][]{2016MNRAS.460.2385W}.
\item We have taken $\approx 50\,{\rm hours}$ of observations using the SCUBA2 instrument on the James Clerk Maxwell Telescope (JCMT) at  a wavelength of $850\,\mu{\rm m}$. These observations will be crucial for identifying which of the many sources we have detected are indeed DSFGs. At present maps have achieved a r.m.s. noise level of $\approx 3\,{\rm mJy}$. 
\item We have observed part of the SuperCLASS field with the Arcminute Microkelvin Imager \citep[AMI;][]{2008MNRAS.391.1545Z} between August 2016 and May 2017, using both the Large Array (LA) and Small Array (SA) which comprise eight $12.8\,{\rm m}$ (ten $3.7\,{\rm m}$) dishes respectively.  AMI operates at a central frequency of $\sim 15\,{\rm GHz}$, with an effective bandwidth of $4\,{\rm GHz}$. With the LA, the primary beam FWHM is $\sim 5.5\,{\rm arcmin}$ and therefore the Northern region of the SuperCLASS field was surveyed using twelve raster mosaics of 20 pointings each. The SA possesses a larger primary beam ($\sim 20\,{\rm arcmin}$) allowing the same survey area to be covered by twelve single pointings. We will use these LA and SA observations jointly to (i) classify the source population at high frequencies and low flux densities, and (ii) search for evidence of the Sunyaev-Zel'dovich (SZ) effect from the constituent clusters of the supercluster. For the AMI-LA data, the typical sensitivity is $32\,\mu{\rm Jy}\,{\rm beam}^{-1}$, and the resolution is $50\times32\,{\rm arcsec}$. The LA source catalogue is discussed in \cite{2018MNRAS.474.5598R}. 
\end{itemize}

\section{Radio Data Products}
\label{sec:products}
Following the independent imaging of {\em e}-MERLIN and \vla data as described in \cref{sec:obs} above, we perform source finding on each of these images, validate the catalogues produced by the source finder, and add a number of flags in order to define samples useful for weak lensing.
\subsection{Definition of DR1 region}
\begin{figure*}
\includegraphics[width=0.975\textwidth]{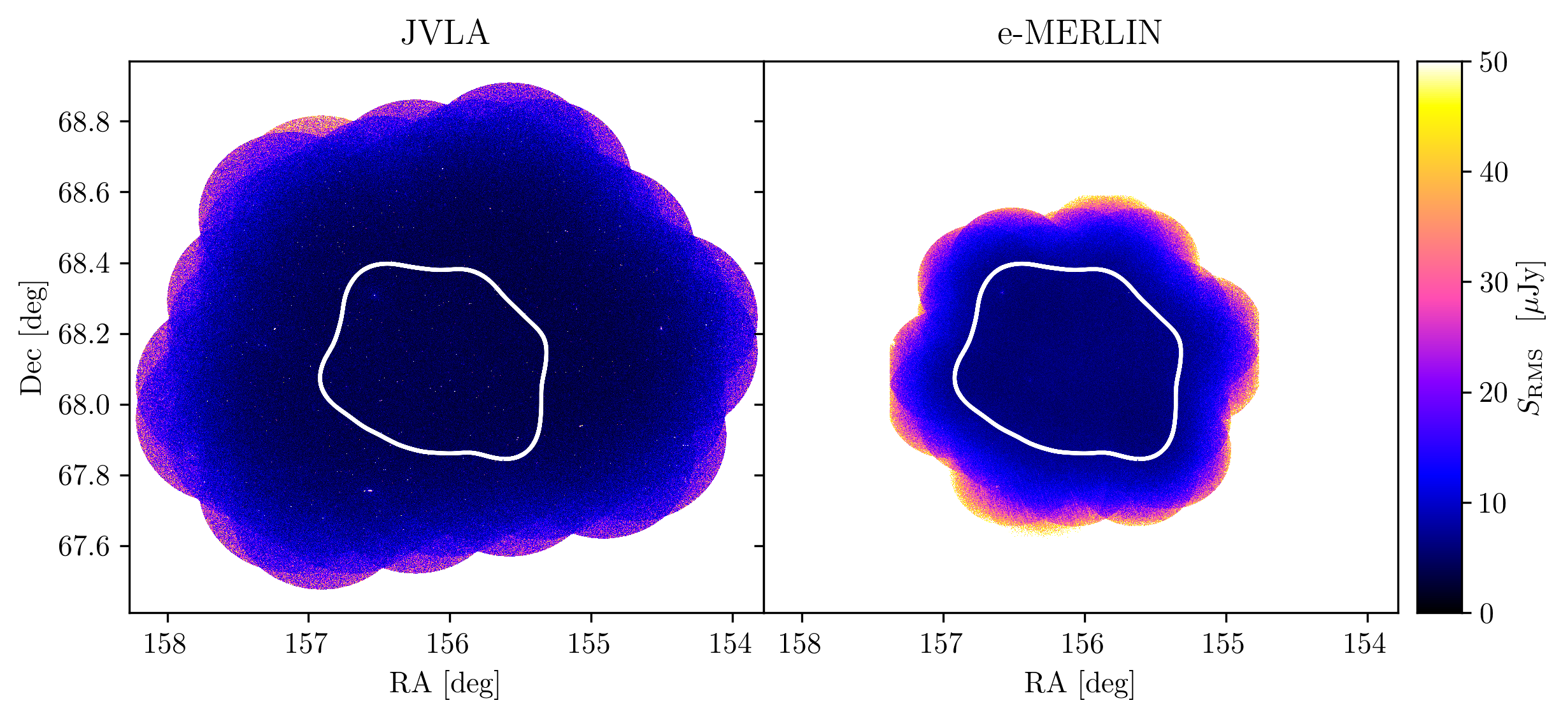}
\caption{\emph{Left}: a map of the r.m.s. noise in the \vla mosaic image for data used for DR1. \emph{Right}: the same for the {\em e}-MERLIN mosaic image. In each case the solid white line denotes the DR1 region, which is the region with noise $< 7\,\mu$Jy in the \emerlin r.m.s map.}
\label{fig:rms_maps}
\end{figure*}

We first define the DR1 region of the SuperCLASS field to be that over which the r.m.s. noise in the mosaiced {\em e}-MERLIN image is approximately constant and below the level of 7$\,\mu{\rm Jy}\,{\rm beam}^{-1}$. From the image mosaics we create the r.m.s. maps shown in \cref{fig:rms_maps}, which show the DR1 region indicated by the solid contour and has an area of $\approx 0.26\degsq$.
\subsection{Released data products}
\label{sec:released}
\begin{figure}
\includegraphics[width=0.5\textwidth]{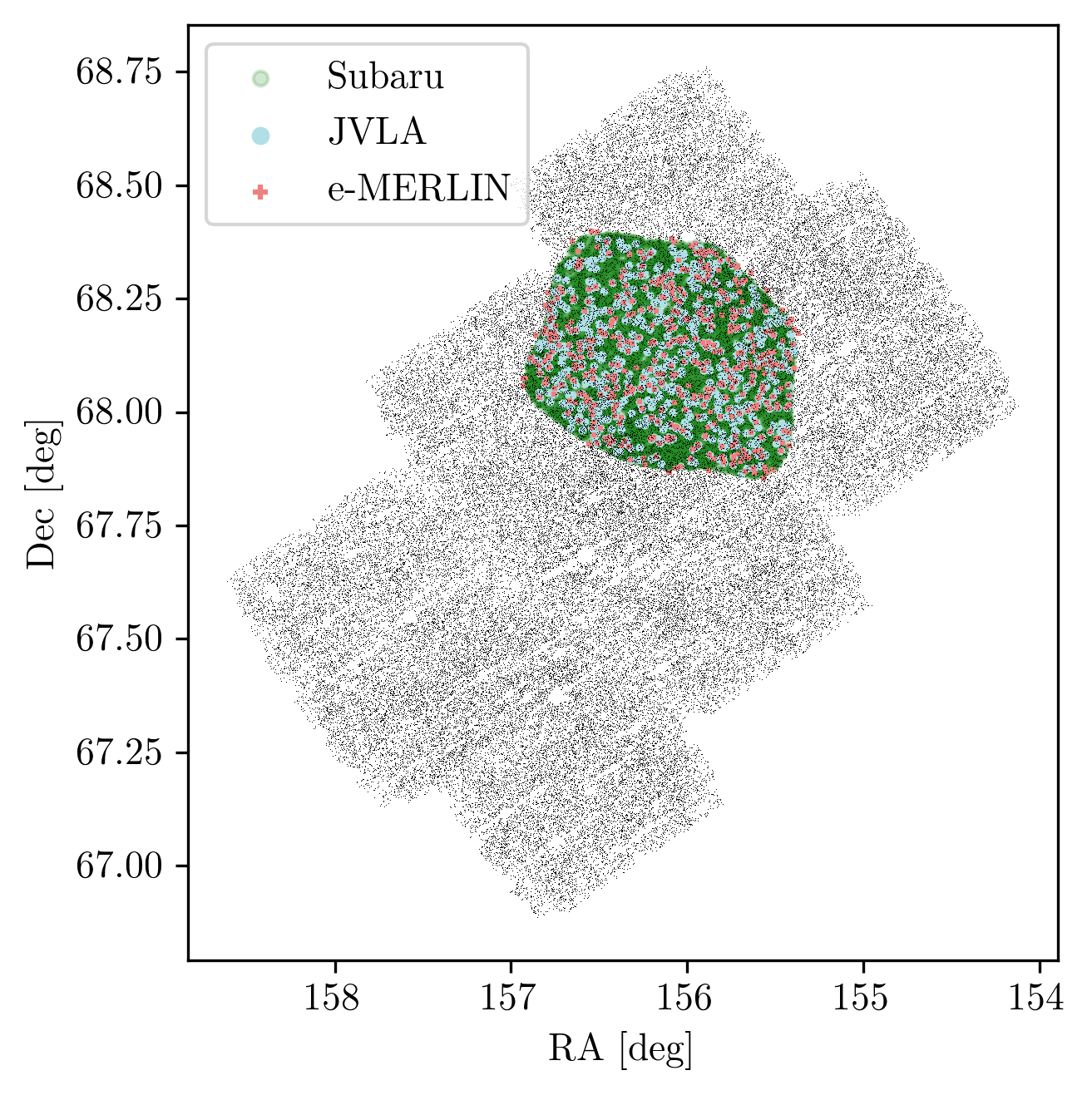}
\caption{Comparison of the DR1 region (shown by the positions of sources as marked in colours: red for \emerlin, blue for \vla and green for Subaru) to the larger full optical region as shown by positions of sources marked by the black points.}
\label{fig:optical_dr1_region}
\end{figure}

We make catalogues available for the \emerlin and \vla DR1 regions. Both of these catalogues contain the columns from the \pybdsf code as `detection' columns, along with columns marking the sources as resolved and having simple morphology. The \vla catalogue also includes columns relating to shape measurements of the sources as described in Paper III. We also provide the Subaru optical catalogue for the full SuperCLASS region (the black points in \cref{fig:optical_dr1_region}). This catalogue contains columns related to the redshift estimation detailed in Paper II and the weak lensing shape measurement detailed in Paper III.\footnote{\ih{\url{ftp://cdsarc.u-strasbg.fr/pub/cats/J/MNRAS/XXX/XXXX}}}
\subsection{Detection of sources}
We carry out source detection using the publicly available {\sc PyBDSF} code \citep{2015ascl.soft02007M}. {\sc PyBDSF} estimates the r.m.s. noise in the image, and finds sources by creating islands of pixels above a specified ``False Detection Rate''. These islands are then fitted with multiple Gaussians, which may be of different sizes, and grouped into sources. For the {\em e}-MERLIN image, the initial run of {\sc PyBDSF} missed a number of resolved, low surface brightness sources which were visible by eye within the image. In order to account for this we smooth the {\em e}-MERLIN image to two scales larger than the original $0.4\,$arcsec beam and run the source finder again (further smoothing with larger scales found no new sources). The union of the catalogues, excluding duplicates, found at the three smoothing scales ($0.4, 0.6, 2.0\,$arcsec) forms our base {\em e}-MERLIN catalogue. The scale at which each source is found is indicated in the {\tt Scale\_flag} catalogue column. For the \vla image, re-running the source finder at larger smoothing scales was found to produce no new catalogue entries, meaning all sources are from the original mosaic image smoothed with a $1.9 \times 1.5\,$arcsec {\tt CLEAN} beam, defining the equivalent \vla catalogue.

\subsection{Preliminary source classifications}
\label{sec:source_classes}
In order to facilitate our science analyses, we add a number of flag columns to the catalogues which pertain to a simple attempt at classification of the sources.

We first classify sources according to the ratio between their integrated and peak fluxes $S_{\rm int}/S_{\rm peak}$. For unresolved sources unaffected by noise, this should be one. However, noise fluctuations in the image may cause scattering around the one-to-one line. In order to remove sources with unphysical values we find the minimum value of $S_{\rm int}/S_{\rm peak} = r_{\rm min}$ and remove all sources with $|1 - S_{\rm int}/S_{\rm peak}| < r_{\rm min}$. For the {\em e}-MERLIN catalogue this leads to the removal of 143 sources, but none are excluded  from the base \vla catalogue. These sources are not just flagged, but are removed entirely from the catalogues.  The number of sources which pass this test are 395 for {\em e}-MERLIN and 887 for \vla.

For the sources which pass this test, the column {\tt Resolved\_flag} has the value {\tt True} when a source meets all three of the criteria:
\begin{align}
\mathtt{DC\_maj} > 0\,, \nonumber\\
\mathtt{DC\_min} > 0\,, \nonumber\\
\mathtt{Maj} > \theta_{\mathrm{Maj}}, \nonumber
\end{align}
where $\theta_{\mathrm{Maj}}$ is the major axis of the {\tt CLEAN} beam for the image.

The column {\tt Simple\_morphology\_flag} has the value {\tt True} for sources which do not show strong visual evidence for having a complex morphology, that is, jet structure in an AGN or a well-resolved irregular star-forming galaxy. We expect sources with a simple morphology to be dominated by star-forming galaxies, and use this set for the shape measurements described in Paper III. In order to create this column we used a private project on the Zooniverse\footnote{\url{https://www.zooniverse.org/}} website to provide us with an initial visual classification of identified sources, with the aim of filtering out multi-source associations which are likely to be AGN. For each detected source, users were presented with cut-outs of the {\em e}-MERLIN mosaic image, the \vla mosaic image, and the Subaru $i^\prime$-band mosaic image, and asked to classify the source as one of:
\begin{enumerate}
\item single source (SFG-like);
\item part of multi-source association (AGN-like);
\item other / uncertain.
\end{enumerate}
A total of five users performed classifications on all of the available sources, with sources being classed as {\tt Simple\_morphology\_flag} sources where they were not classified as type (ii) or type (iii) sources by a majority of users. 

One of the science goals of the SuperCLASS project is to develop a classification scheme for the detected radio sources based both on morphology and photometric characteristics which is on-going \citep[see][]{superclass2}. For the purposes of this paper we merely attempt a crude preliminary split of the detected radio sources by {\tt Simple\_morphology\_flag=True}. For weak lensing analyses it is helpful to separate the source population into these two broad categories since they would need to be modelled somewhat differently - note that any future SKA weak lensing survey, being much deeper than any survey done so far, will be dominated by star-forming galaxies with simple morphology (compared to jet dominated AGN). Therefore, the reader who has a more sophisticated understanding of extragalactic object classification should see this as mainly an exercise to assist the preliminary weak lensing analyses (see Paper III), rather than for it to be used for probing the nature of galaxy evolution. Nonetheless, we would hope that it would not lead to entirely misleading results if used, with caution, for that purpose.

\begin{figure*}
\includegraphics[width=0.975\textwidth]{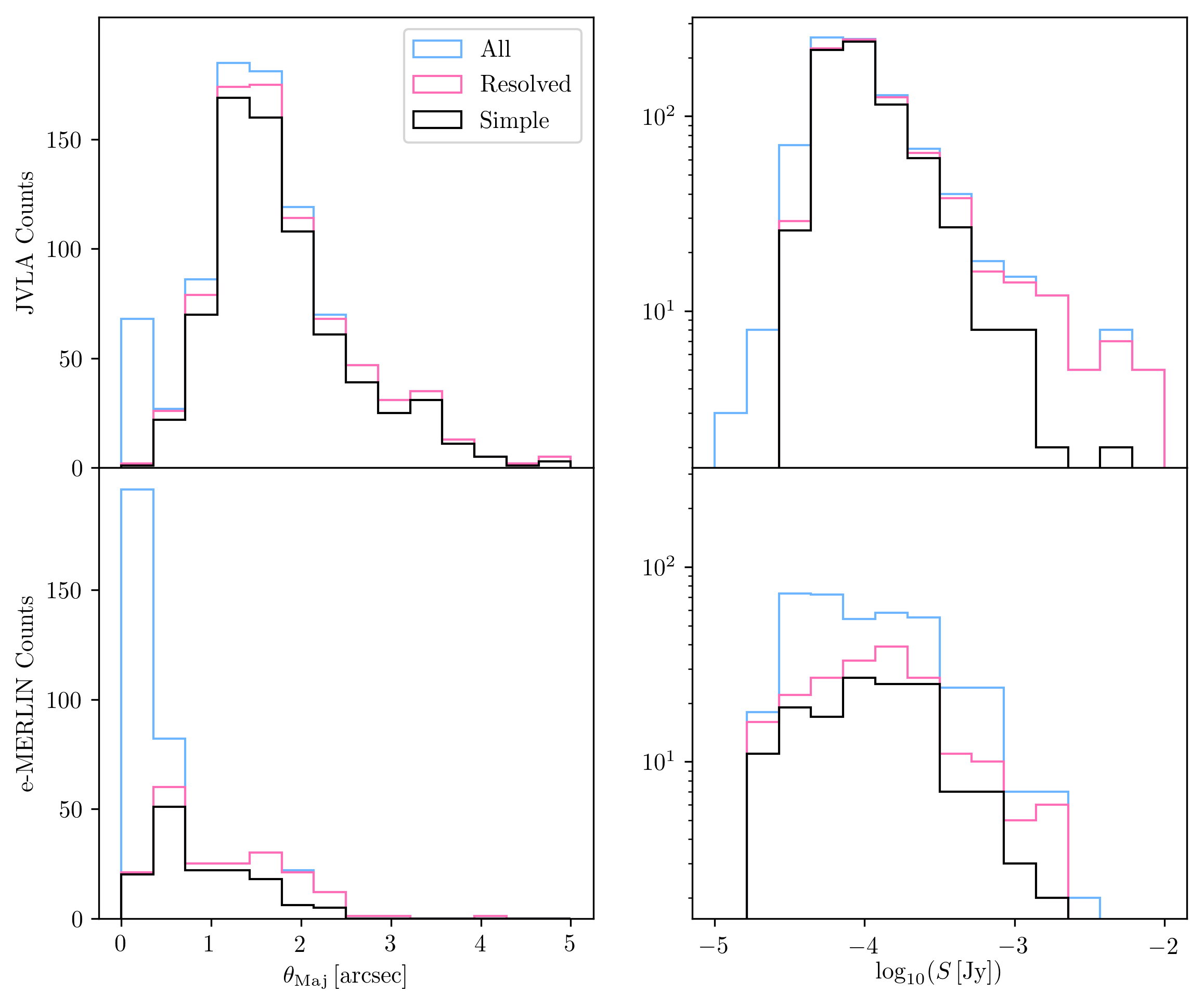}
\caption{Properties of \vla (\emph{upper} panels) and {\em e}-MERLIN (\emph{lower} panels) detected sources. \emph{Left:} the major axis best-fitting Gaussian FWHM as reported by \textsc{PyBDSF}. \emph{Right:} the flux density of the detected sources. In each case we present three histograms: (i) including all sources (cyan), (ii) including only those sources that are resolved (pink) and (iii) including only those sources that have simple morphologies (black). See the main text for precise definitions of these latter two subsamples. See \cref{sec:source_sizes} for a discussion of the reliability of these size measurements.}
\label{fig:fluxsize}
\end{figure*}

\begin{figure*}
\includegraphics[width=0.95\textwidth]{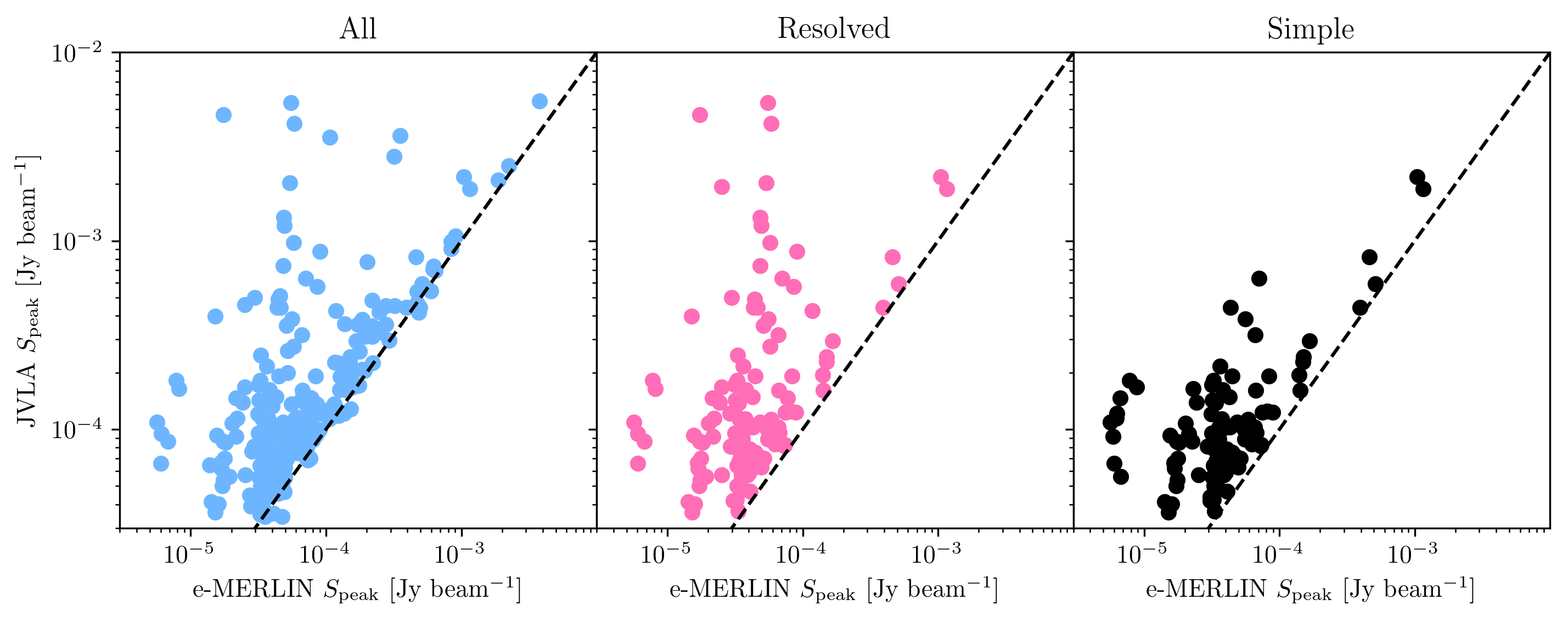}
\caption{Comparison between the peak {\em e}-MERLIN fluxes and those from the \vla for the cross-matched detections, those which are resolved in both and those deemed ``simple'' under our preliminary source classification. Once we concentrate only on sources which are resolved the \vla flux is larger than the {\em e}-MERLIN flux as one would expect -- the longer baselines of {\em e}-MERLIN resolve out much of the extended flux on $>1\,$arcsec scales.}
\label{fig:flux_ratios}
\end{figure*}

Numbers of sources in the {\em e}-MERLIN and \vla SuperCLASS DR1 catalogues meeting these criteria are shown in \cref{tab:source_numbers} and summary plots of the fluxes and sizes are shown in \cref{fig:fluxsize}. In addition, we show the behaviour of the peak flux ratios between {\em e}-MERLIN and \vla in \cref{fig:flux_ratios}.

\begin{table}
\caption{Numbers of entries in the SuperCLASS DR1 radio catalogues, when sources from flagged columns are successively excluded.}
\label{tab:source_numbers}
\begin{center}
\begin{tabular}{lcc}
\hline
 & {\em e}-MERLIN & \vla \\
\hline
Full DR1 & 395 & 887 \\
\texttt{Resolved\_flag}$=$\texttt{True} & 197 & 789 \\
\texttt{Simple\_morphology\_flag}$=$\texttt{True} & 144 & 710 \\
\hline
\end{tabular}
\end{center}
\end{table}

\begin{figure*}
\centerline{\includegraphics[width=0.8\textwidth]{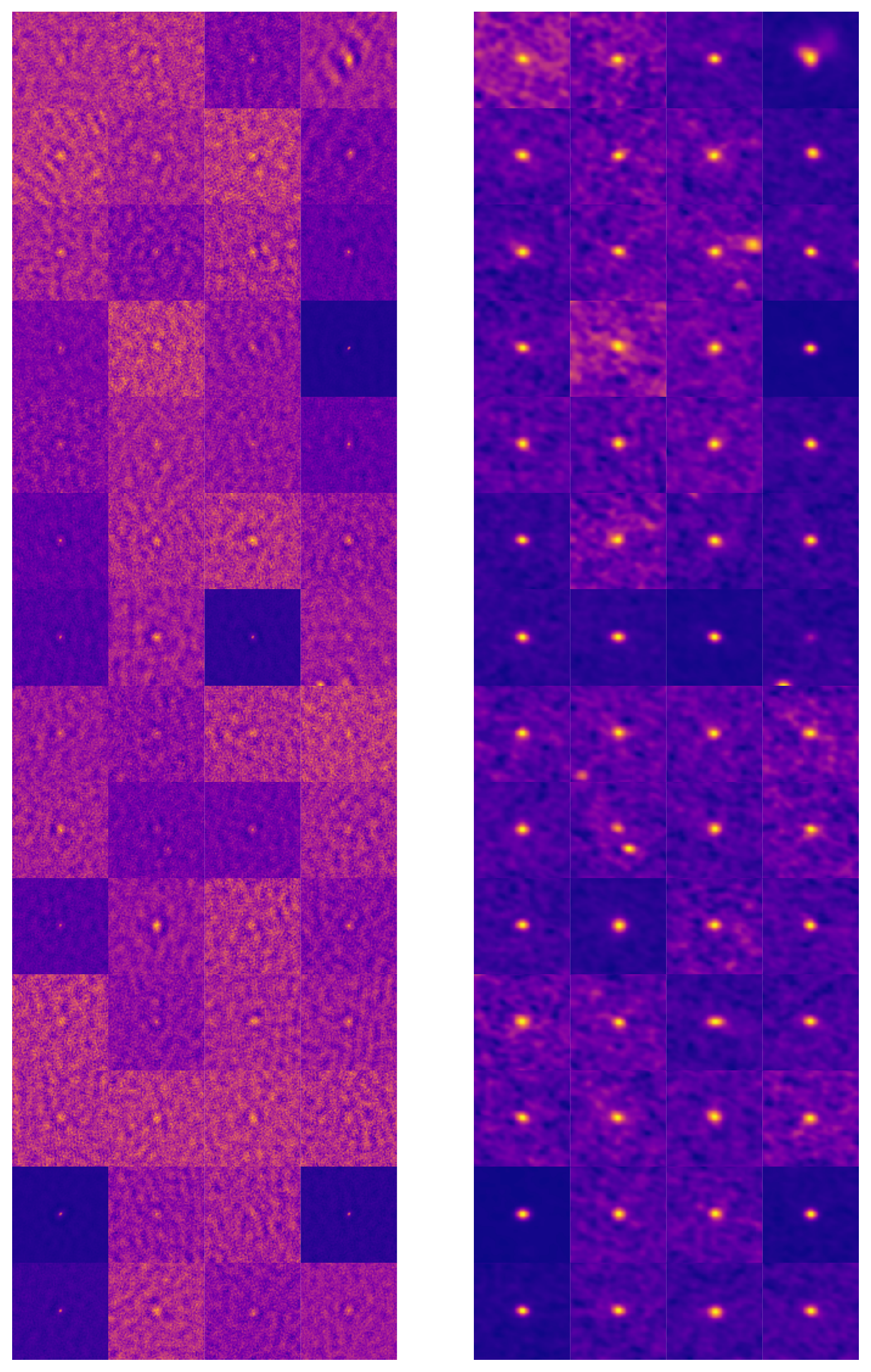}}
\caption{The 56 sources with flux $>100\,\mu$Jy which are detected and resolved by both {\em e}-MERLIN and \vla, showing the data quality available from our observations and the scales probed by the two different observatories. On the \emph{left} are the {\em e}-MERLIN images and on the \emph{right} are the \vla images. Each thumbnail is ten arcsec across and colour scales are normalised individually for each one to emphasise the morphology of sources.}
\label{fig:thumbnails}
\end{figure*}

From the \emerlin data we select 56 high-confidence sources, requiring them to have \texttt{Resolved\_flag=True} in both \emerlin and \vla catalogues, \texttt{Total\_flux}$\,> 100\,\mu$Jy in both \emerlin and \vla, and \texttt{Simple\_morphology\_flag=True} indicating they have non-complex morphology (our proxy for being star-forming galaxies). Thumbnails of these sources are shown in \cref{fig:thumbnails}, while the associated matching information and Source IDs are provided in \cref{tab:thumbnails}.

\begin{table*}
\caption{Fluxes and separations in {\em e}-MERLIN and \vla images of the 56 sources shown in thumbnails in \cref{fig:thumbnails}. \ih{The full catalogues containing these sources is available at the CDS VizieR service \url{ftp://cdsarc.u-strasbg.fr/pub/cats/J/MNRAS/XXX/XXXX}.}}
\label{tab:thumbnails}
\begin{tabular}{lccccc}
\hline
Source ID {\em e}-MERLIN & Peak Flux & Peak Flux & Total Flux & Total Flux & Separation \\
 & {\em e}-MERLIN  $\mathrm{[Jy/beam]}$ & \vla $\mathrm{[Jy/beam]}$ & {\em e}-MERLIN  $\mathrm{[Jy]}$ & \vla $\mathrm{[Jy]}$ & $\mathrm{[arcsec]}$ \\
 \hline
SCL-EM-J102458.43+675300.37 & 0.05 & 0.06 & 0.26 & 0.10 & 0.24 \\
SCL-EM-J102416.01+675237.81 & 0.04 & 0.07 & 0.22 & 0.11 & 0.15 \\
SCL-EM-J102409.08+675247.19 & 0.08 & 0.19 & 0.18 & 0.26 & 0.09 \\
SCL-EM-J102226.97+675447.27 & 0.07 & 0.63 & 1.62 & 1.83 & 0.46 \\
SCL-EM-J102545.40+675636.41 & 0.04 & 0.11 & 0.28 & 0.20 & 0.04 \\
$\vdots$ & $\vdots$ & $\vdots$ & $\vdots$ & $\vdots$ & $\vdots$  \\
 SCL-EM-J102314.77+682004.35 & 0.07 & 0.16 & 0.22 & 0.23 & 0.06\\
\hline
\end{tabular}
\end{table*}

\subsection{Forced source photometry}
\label{sec:forced}
As the \vla synthesised beam (the interferometer PSF or `dirty beam') is a better match to the expected source morphology than the \emerlin one it achieves a better blind detection rate of objects, even though both {\em e}-MERLIN and \vla surveys have a similar noise level in terms of ${\rm Jy}\,{\rm beam}^{-1}$. For the \vla source positions where no corresponding {\em e}-MERLIN source was found, a non-blind forced fit can be made in the {\em e}-MERLIN maps using this extra information of the positions of the \vla sources. All or some of the source flux can be hidden in the {\em e}-MERLIN images because the surface brightness is close to or below the noise level, but a simple application of aperture photometry should be able to recover the flux. This is also a robust alternative to source fitting for larger diffuse sources based on a model, such as a S\'{e}rsic or Gaussian profile, as it is little affected by the source shape.

\begin{figure}
\includegraphics[width=0.475\textwidth]{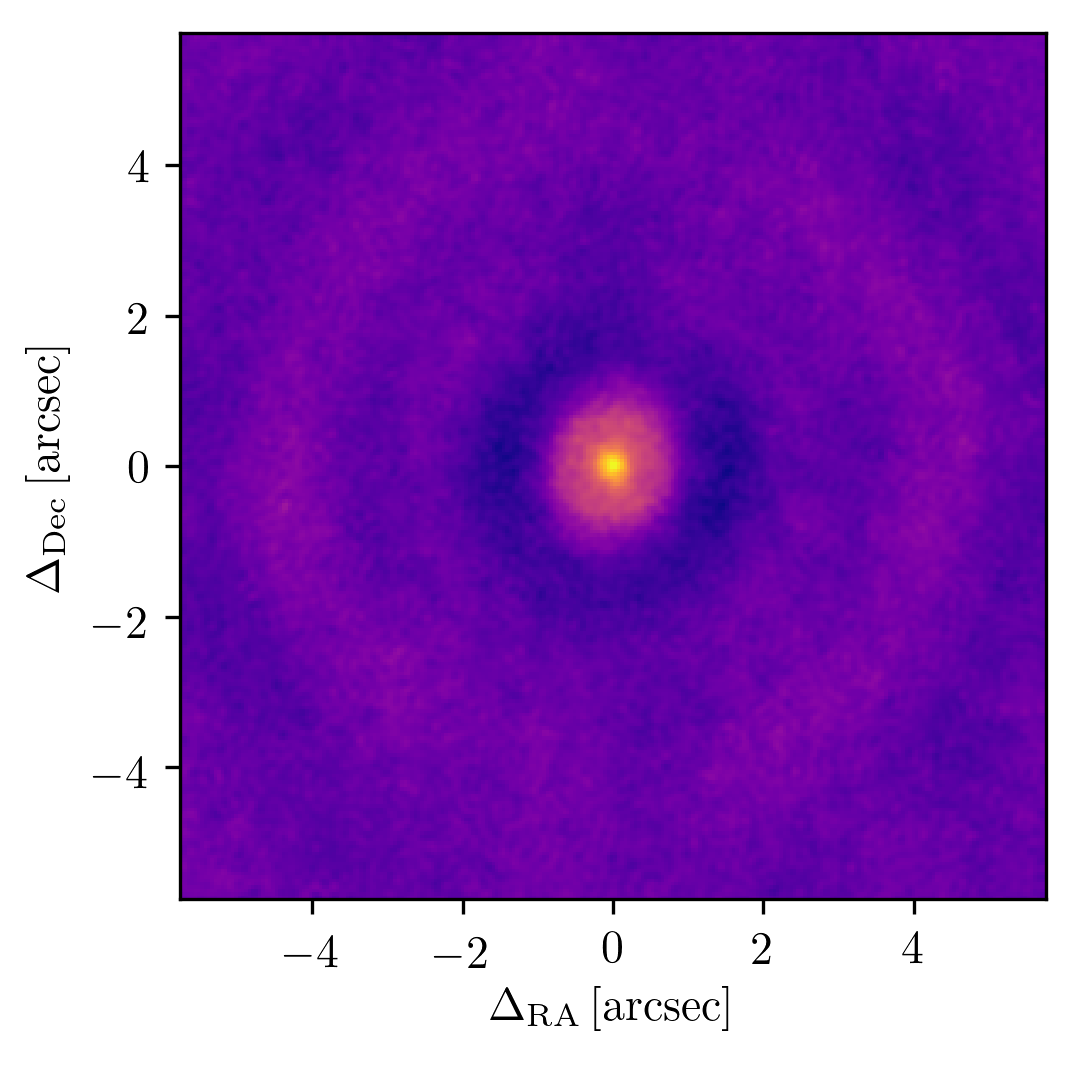}
\caption{\emph{e}-MERLIN image data, stacked at locations where a detection of a source was made in the VLA image, but not the \emph{e}-MERLIN image. Undetected flux from these sources adds coherently with the shape of the dirty beam, showing that information is contained within the \emerlin data from the sources which were undetected in the \emerlin-only image.}
\label{fig:emerlin_stack}
\end{figure}

A simple aperture photometry script was implemented in python which takes $256 \times 256$ cut-outs centred on the \vla positions. The image was converted to units of Jy per pixel using the solid angle of the \clean beam given in the fits header. To estimate the source flux the signal was integrated within a 1 arcsec radius aperture and subtracted from the median value in an outer annulus of the same area, multiplied by the number of pixels in the central circular aperture. Errors were determined by using the standard deviation of repeated measurements in 99 positions over the cut-out. A stack of the cut-outs is shown in \cref{fig:emerlin_stack}, with the flux of the individually undetected sources now clearly visible and following the expected pattern of the \emerlin dirty beam PSF.

From the \pybdsf catalogues, 246 of the 395 sources detected in the \emerlin image have matching \vla sources in a 1 arcsec search radius. For the 641 un-matched sources the aperture photometry measurements in the \emerlin image do in fact show significant positive flux measurements. For the matched sources, a comparison between aperture flux and \pybdsf measurements shows that those with aperture fluxes $\gtrsim 0.15\,$mJy have matches in the modelled catalogue.

As a sanity check the aperture photometry was repeated but using positions offset by 10 arcsec in a random direction from the \vla position. The correlated addition of flux then disappears leaving a symmetric distribution of measured aperture fluxes around zero as expected for just noise. This strongly suggests the bias above zero flux is due to real source flux seen in {\em e}-MERLIN due to the \vla sources.

\subsection{Optical matches}
\begin{figure}
\includegraphics[width=0.5\textwidth]{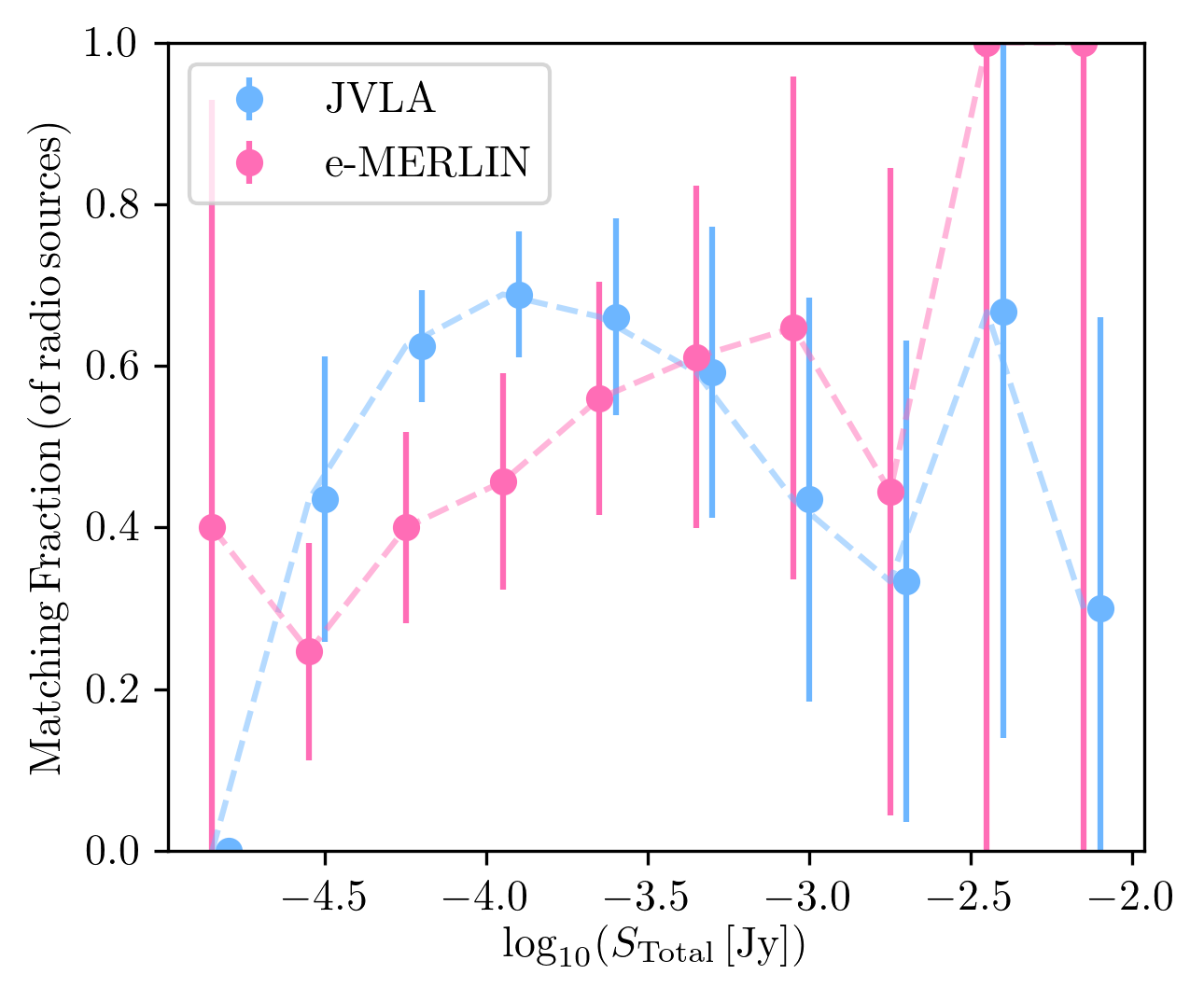}
\caption{Matching fractions of radio sources in \vla and {\em e}-MERLIN catalogues which have a Subaru source within a matching radius of 1 arcsec, as a function of their radio flux.}
\label{fig:radio_optical_flux}
\end{figure}

We also cross-match the {\em e}-MERLIN and \vla catalogues with our Subarau $i^\prime$-band weak lensing shape catalogue (as described in \cref{sec:subaru}), again using a matching radius of 1 arcsec, within the DR1 area. These sources are shown in the context of the full Subaru optical coverage (black points) in \cref{fig:optical_dr1_region}. \cref{fig:radio_optical_flux} shows the fraction of radio sources, by flux, which have matches, with this being between $\sim1/3$ and $\sim2/3$, dependent on the flux range, for both {\em e}-MERLIN and \vla sources. This is broadly consistent with the matching fractions of L-band sources previously found in studies of the Hubble Deep Field-North by \cite{Chapman2003, Barger2007}.

\section{Radio source properties}
\label{sec:science}

\subsection{Source counts}
\begin{figure}
\includegraphics[width=0.475\textwidth]{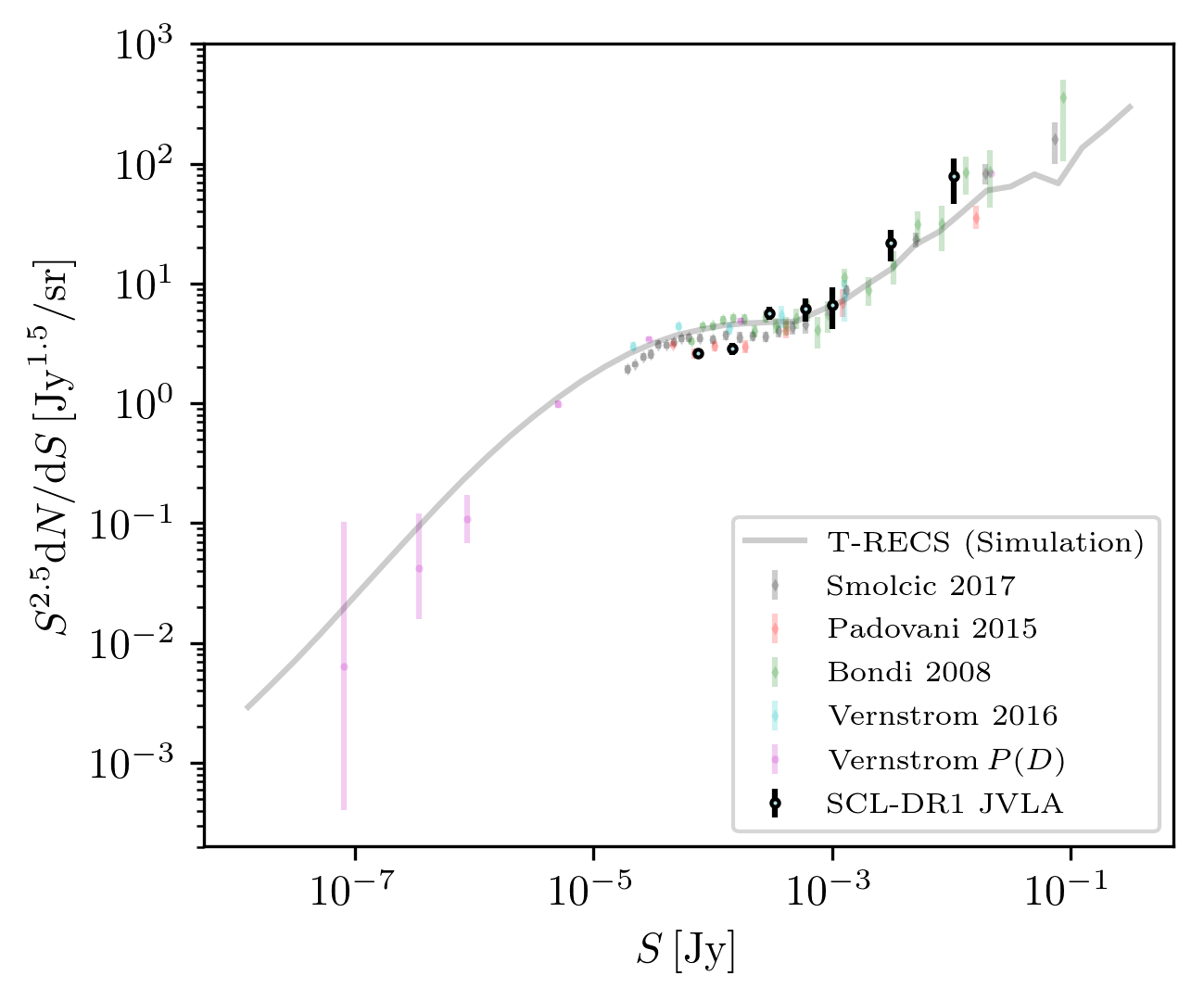}
\caption{Euclidean-normalised source counts for the SuperCLASS \vla DR1 region, shown alongside other L-band flux counts from the literature \citep{2008ApJ...681.1129B,2014MNRAS.440.2791V,2015MNRAS.452.1263P,2016MNRAS.462.2934V,2017A&A...602A...2S}, and the T-RECS simulation source counts at the same frequency.}
\label{fig:dnds}
\end{figure}

As well as the pure flux counts shown in \cref{fig:fluxsize}, we also convert the \vla catalogue source counts to a Euclidean-normalised differential source count in \cref{fig:dnds} and compare to counts from the literature. We take differential counts of the sources (all those detected) in logarithmic bins and multiply with the value of the flux bin centre to the power of $5/2$ and divide by the area of the DR1 region in Steradians and the difference in bin central flux.

We see from \cref{fig:dnds} \ih{that our estimates of the differential source count are broadly compatible with those already in the literature over the relevant range of flux density. However, we caution against taking seriously any discrepancies since we have not attempted to model the completeness of our survey in this analysis. For this reason we have not published specific values for the differential source counts.}

\subsection{Source sizes}
\label{sec:source_sizes}
We also present the source sizes (in terms of the FWHM of the best-fitting Gaussian, including deconvolution of the \clean restoring beam, as determined by {\sc PyBDSF}) for the \vla catalogue in \cref{fig:fluxsize}. However, we have reason to believe these measured source sizes are as much a function of the instruments involved as the intrinsic source sizes. Distributions for both telescopes peak around the sizes of the \clean restoring beam, even where the `resolved' criteria of \cref{sec:source_classes} are applied. The procedure of fitting Gaussian profiles to \clean image reconstructions -- where Gaussians of the size of the restoring beam are added back to the residual image -- is not one which can be expected to return correct morphological information for true sources with intensity profiles which are different to that of a Gaussian. Of the two true dirty beam PSFs involved, the \vla PSF is large ($1.9 \times 1.5\,$arcsec) compared to the expected $\sim 1\,$arcsec size of the star-forming galaxy sources \citep[see e.g.][]{2008MNRAS.385..893B}, and the \emerlin PSF, which has significant negative sidelobes, will `resolve out' much of the emission from diffuse profiles with shallow wings.
   
\begin{figure*}
\includegraphics[width=\textwidth]{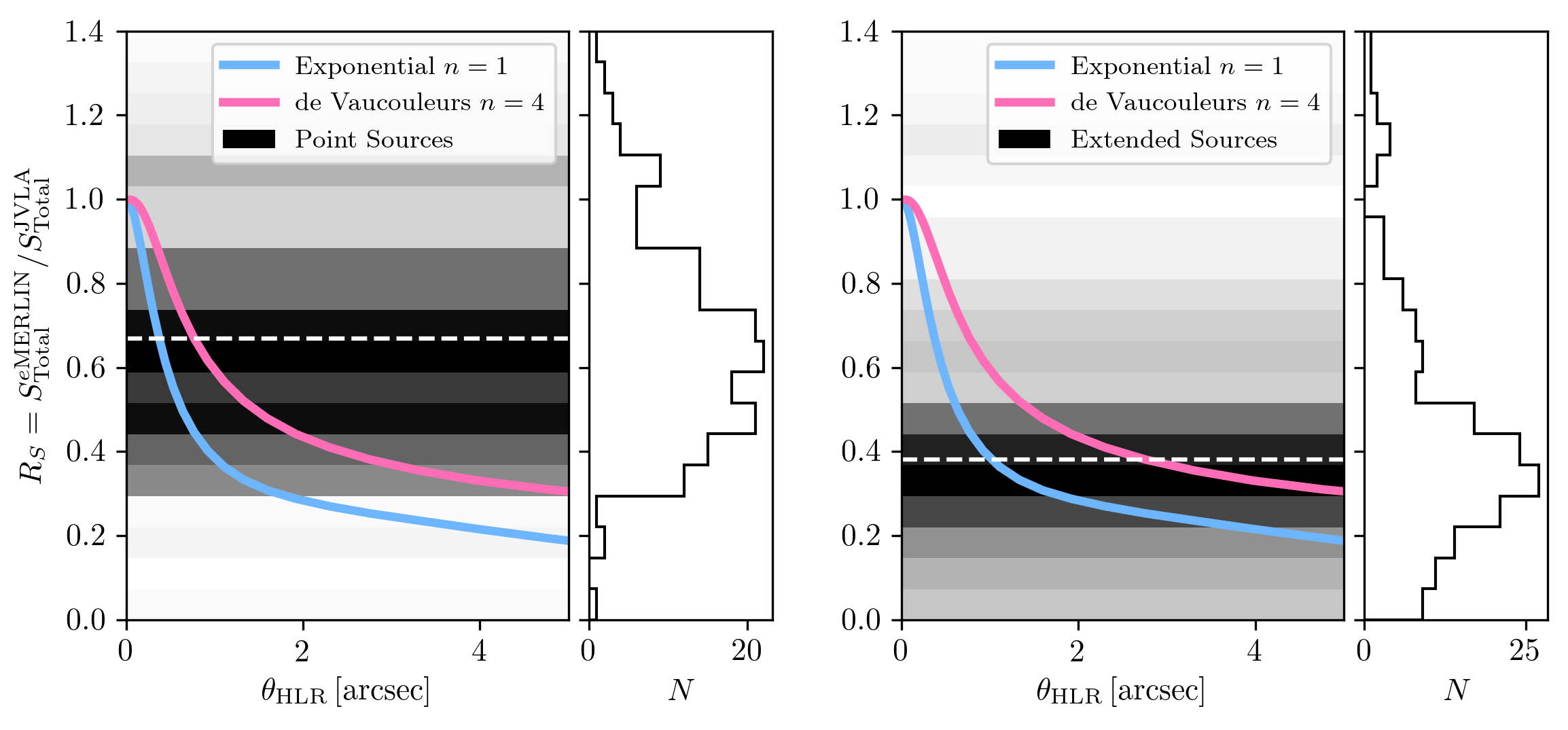}
\caption{\emph{e}-MERLIN to VLA flux ratios as measured for simulated point sources (\emph{left}) and extended sources (Exponential profiles, \emph{right}). Red and blue lines represent the expectation for different \sersic profiles of differing half light radii. The flux ratio constrains the degenerate combination of source size and \sersic index. Horizontal bars represent histogram counts of measured source flux ratios, with the median shown by the white dashed line. These plots should be compared with those below in \cref{fig:flux_ratio_real}, which show the degenerate combination of \sersic index and half light radii of the observed sources is more consistent with them being extended sources than point sources.}
\label{fig:flux_ratio_sim}
\end{figure*}

To further quantify this effect we investigate a robust, model-independent quantity related to the extent of the detected sources: the ratio of total measured source fluxes between \emerlin and \vla images which we refer to as $R_S$. This quantity is constructed by measuring the total flux within apertures as described in \cref{sec:forced} at the locations of sources detected in the \vla images. Using the $uv$-coverage of both telescopes, we simulate the flux recovered by each for sources parameterised with a \sersic intensity profile as a function of radius:
\begin{equation}
    I(\theta) = I_{\rm HLR} \exp \left\lbrace -b_n \left[ \left( \frac{\theta}{\theta_{\rm HLR}} \right)^{1/n} - 1 \right] \right\rbrace,
\end{equation}
where $\theta_{HLR}$ is the half-light radius of the profile, $n$ is the \sersic index parameter which sets how steeply the intensity profile decreases and $b_n \approx 2n - 1/3$. \Cref{fig:flux_ratio_sim} shows this quantity from the simple simulations as the solid lines, one for an $n=1$ \sersic profile (corresponding to an exponential profile) and one for an $n=4$ \sersic profile (corresponding to the much `peakier' de Vaucouleurs profile, shown as an example for more compact emission, though we do not expect radio star-forming galaxies to have this profile). The \emerlin to \vla total flux ratio is shown as a function of the half-light radius of the simulated source. As can be seen, $R_S$ goes to one as expected for sources with no spatial extent ($\theta_{\rm HLR} \rightarrow 0$) which are unresolved by both telescopes. However, as the $\theta_{\rm HLR}$ of both source profiles is increased $R_S$ decreases significantly, mostly due to the `resolving out' effect of the interferometers, whereby diffuse flux is not captured due to a lack of smaller spacings between the interferometer elements. For model sources with these sizes and \sersic indices, this `resolving out' effect is more important for \emerlin than it is for the \vla.

To confirm these simple simulations we also run our full simulation pipeline (as created for the weak lensing science and described in Paper III), consisting of a source flux-size population model from the T-RECS simulation \citep{2019MNRAS.482....2B}, $uv$-data creation and imaging, and measurement of aperture fluxes. We do this with two sky models, one in which all sources are point sources (\Cref{fig:flux_ratio_sim}, left panel) and one in which sources follow the input T-RECS size distribution but have exponential intensity profiles (\Cref{fig:flux_ratio_sim}, right panel). Histograms of $R_S$ measured in these two simulations are shown as the greyscale histogram bars, and the median is shown by the white dashed line. As can be seen, the values of $R_S$ measured for the exponential sources is consistent with the true distribution of $n=1$ profiles with a median half-light radius $\sim 1\,$arcsec, and the point sources (which are more affected by correlated noise fluctuations in the imaging process) are marginally consistent with the expected $R_S$ of one.

\begin{figure*}
\includegraphics[width=\textwidth]{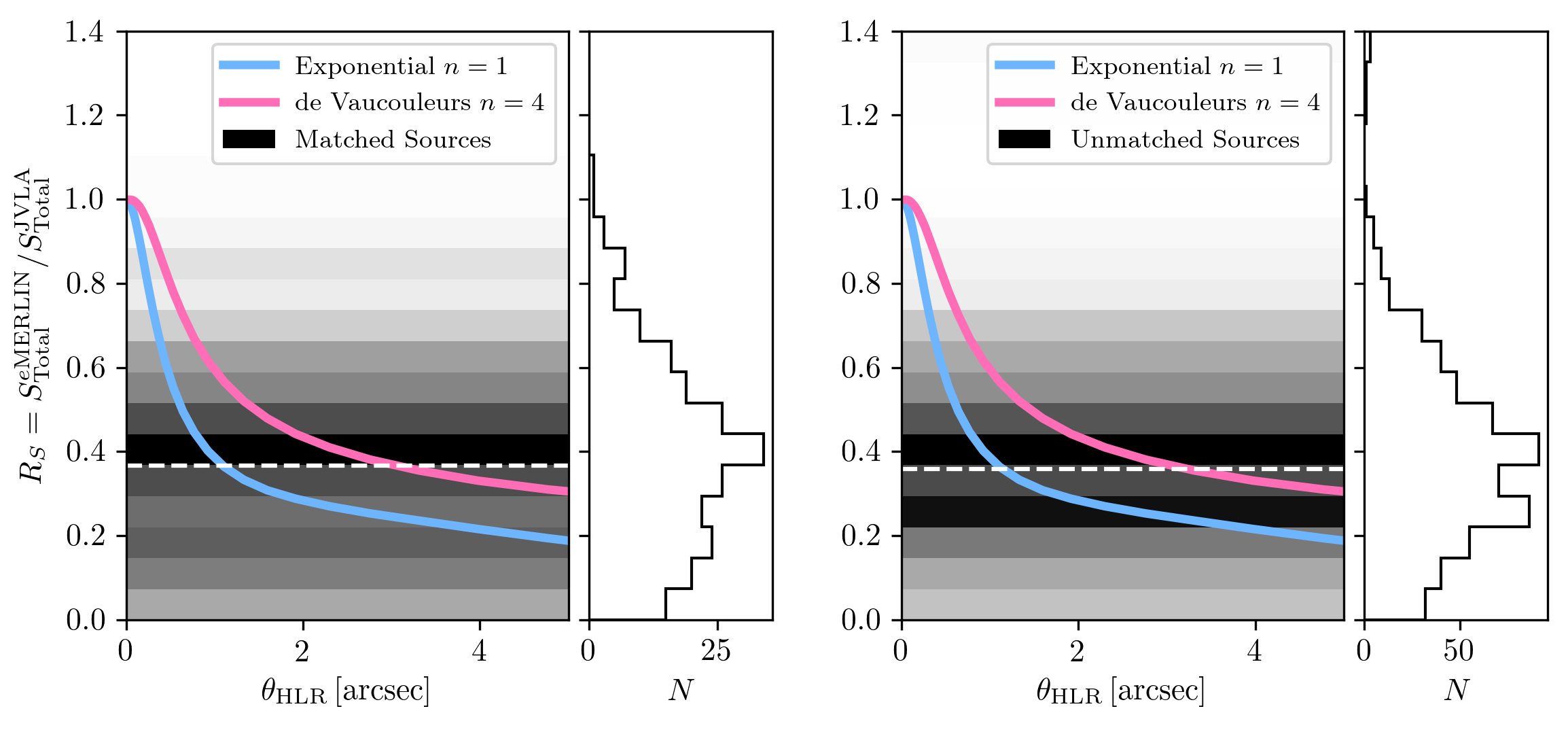}
\caption{Same as \cref{fig:flux_ratio_sim} but for the real DR1 catalogue. \emph{Left panel:} the measured ratios for sources which are detected independently in both the \emph{e}-MERLIN and VLA images. \emph{Right panel:} measured ratios for those sources that are detected in the VLA image only, with forced photometry at the corresponding location in the \emph{e}-MERLIN image. The measured $R_S$ for both un-matched and matched sources is consistent with them being extended, rather than point sources but cannot measure independently their \sersic index or radius.}
\label{fig:flux_ratio_real}
\end{figure*}

We compare these results to the same quantity $R_S$ measured in apertures around \vla positions in our DR1 images in \cref{fig:flux_ratio_real}, both for sources which have matches between the \emerlin and \vla images on the left and those which do not on the right. The measured values are again shown as the greyscale histograms with median indicated by the white dashed lines. As can be seen these measurements constrain the degenerate combination of \sersic index and half-light radius for this population, with the profiles consistent with the $\sim 1 \,$arcsec exponentials expected for star-forming galaxies at these flux densities, as well as de Vaucouleurs profiles with sizes up to $5\,$arcsec and exponential sources with smaller sizes between $0.5-1\,$arcsec.

\subsection{Spectral properties}
An initial investigation of the spectral properties of radio sources in this field was performed by \cite{2016MNRAS.462..917R}. With the new \vla data presented in this work, we now have two highly sensitive catalogues with which to further investigate the spectral index distribution.

The $325\,$MHz SCG325 catalogue contains approximately 213 sources in the region covered by our \vla catalogue. Of these, we find 205 GMRT sources with counterparts at 1.5~GHz. Given the large difference in resolution (13~arcsec at 325~MHz compared to $\approx$2~arcsec at 1.5~GHz), many GMRT sources become resolved into multiple sources in our \vla catalogue. We summed the flux densities of all \vla sources corresponding to each GMRT source when deriving our spectral index distribution, and applied a correction factor 0.972 to ensure consistent flux density scaling between the \vla catalogue (which uses the flux scale of \citealt{2013ApJS..204...19P}) and the SCG325 catalogue (which uses the \citealt{2012MNRAS.423L..30S} scale). 

\begin{figure}
\centerline{
\includegraphics[width=0.475\textwidth]{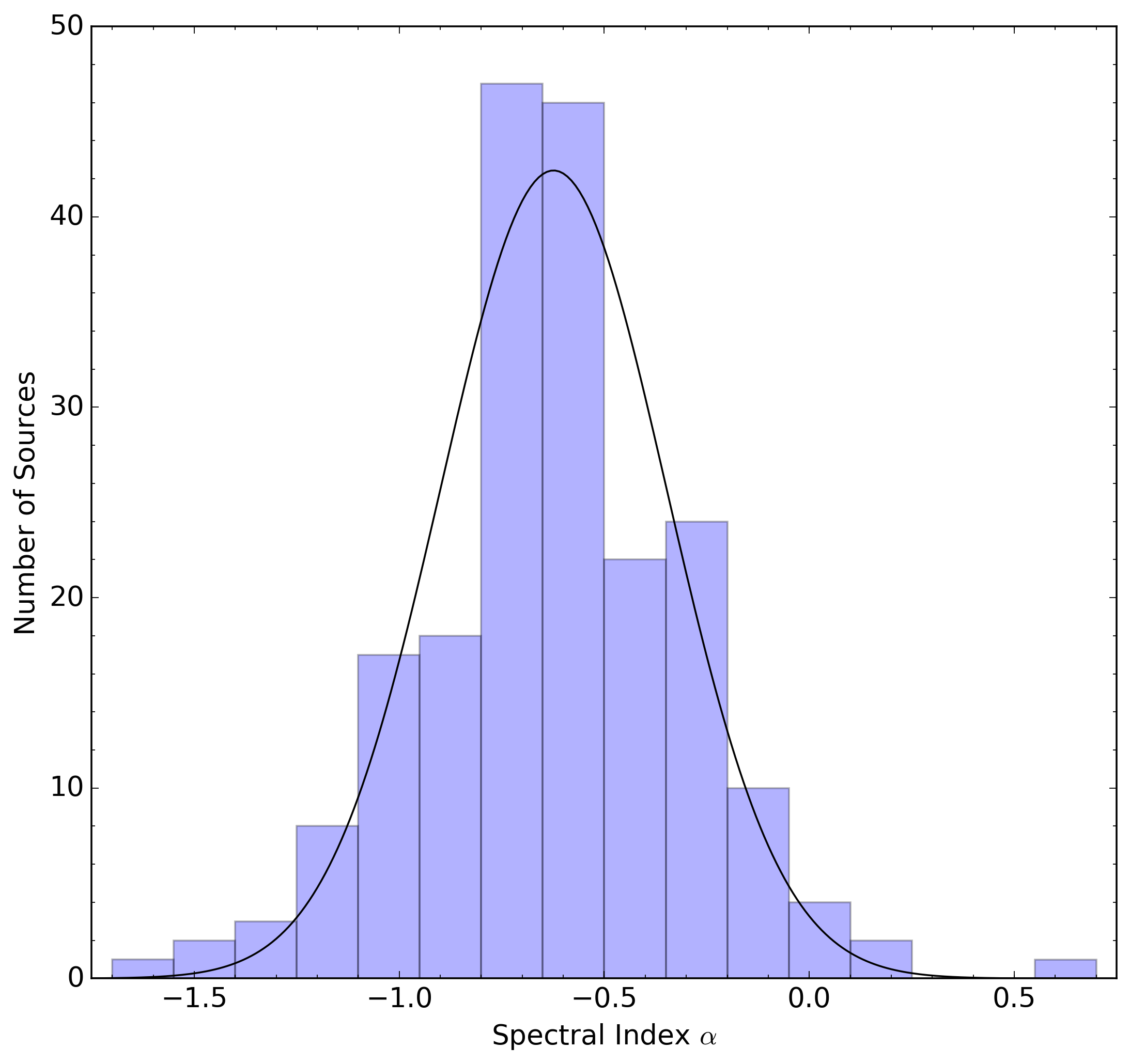}}
\caption{Histogram of the spectral index $(\alpha)$ distribution  for the 205 sources matched between the SCG325 catalogue and our \vla catalogue. The curve denotes the best-fit Gaussian, centred on $\alpha=-0.62$, with $\sigma=0.28$.}
\label{fig:vla_gmrt_alpha}
\end{figure}

We present a histogram of the spectral index distribution in Fig.~\ref{fig:vla_gmrt_alpha}. Our distribution appears reasonably well-described by a single Gaussian centred on $\alpha=-0.62$, although the population exhibits a relatively high standard deviation of $\sigma=0.28$.

Four sources in our cross-matched catalogue\footnote{These are SCL-JV-J102330.18+681211.47, SCL-JV-J102501.59+681918.63, SCL-JV-J102531.59+680853.13 and SCL-JV-J102612.19+680827.02} exhibit positive spectra between 325~MHz and 1.5~GHz. All are compact at the resolution of both the GMRT and the \vla.

\section{Summary and conclusions}
\label{sec:sumcon}
We have presented the Data Release 1 (DR1) data set from the SuperCLASS survey. This data set consists of multi-wavelength observations including in the optical, near-infrared, sub-mm, and at high (GHz) and low (MHz) radio frequencies, all covering a $\sim 1\,\degsq$ field in the Northern sky containing a massive supercluster at a redshift of $\sim0.2$. The aim of these observations is to detect and characterise, in terms of their spectra and morphology, $10^4 - 10^5$ star-forming galaxies covering a redshift range up to $z \sim 1$. This will allow us to: use the statistical properties of their observed shapes to measure a weak lensing shear signal from the intervening dark and baryonic mass along the line of sight; and to understand their formation and evolution through access to their star formation and Active Galactic Nuclei (AGN) activity unobscured by dust absorption.

In this paper we have described the acquisition, reduction and initial source catalogue creation of the DR1 data set. With these catalogues in hand we refer the readers to \cite{superclass2} (Paper II) for further details on studies of the sources' redshifts and matched properties between radio, optical and near-infrared observations. Similarly, \citet{superclass3} (Paper III) describes the use of the source catalogues and images to measure simple morphological properties -- the ellipticities -- of the sources in order to infer the gravitational lensing effect of the super-cluster mass, both in the optical and radio data alone and together in cross-correlation.

We have shown that the flux distribution of our source catalogues, constructed using the \vla mosaic image as a detection image, is consistent with previous studies at similar $\mu$Jy depths at $1.4\,$GHz radio L-band. We have presented catalogues of the properties of these sources as measured by the \textsc{PyBDSF} source finding algorithm. Cross-matching the positions of these sources (and those of sources detected in the \emerlin image) with the positions of sources detected in the all-band Subaru HSC image yielded matching fractions between $1/3$ and $2/3$ of the radio sources, again consistent with expectations of previous studies. Paper II uses this data to fit Spectral Energy Distribution templates to these matched sources and classify them according to AGN and star forming activity. We have also measured spectral indices between the radio L-band sources and their matches in the lower radio frequency catalogue from our GMRT data, again finding good agreement with the literature.

As well as the fluxes of the detected sources, we have also conducted preliminary investigations into their morphology, in particular constraining the \emerlin to \vla total flux ratio $R_S$ of the sources. We have argued that this quantity is more robust to the observational effects of radio interferometers than fitting parametrised Gaussian profiles, in particular where lack of short spacings `resolves out' diffuse flux. By comparing with simulations we have used $R_S$ to constrain the degenerate combination of source \sersic index and size across the population of sources in our \emerlin and \vla catalogues, finding consistency with (but not strong constraints on) the sizes of $\sim 1\,$arcsec with an exponential profile slope. Better constraints on the source morphology are a key goal of the final data release of SuperCLASS, and we intend to enable this through combination of the data from both \emerlin and \vla together in the native $uv$ plane (see the extended discussion in section 6 of Paper III).

We have not discussed the redshift distribution of sources here, instead deferring to the detailed study of Paper II, which finds a prominent over density of sources at $z\sim0.2$ as expected from historical observations of Abell clusters in the region.

The data and catalogues presented here will be extended in the full SuperCLASS data release to include approximately three times the area of L-band radio imaging, greatly increasing the statistical power available to us in constraining both the weak lensing signal and galaxy formation and evolution histories. This data set will be unique in the pre-SKA era in allowing access to small angular scales $\lesssim 1\,$arcsec through its inclusion of data from \emerlin. We therefore expect it to be highly valuable in studying the source populations for large scale cosmology and galaxy evolution surveys with SKA in the next decade. Surveys with the full SKA will be large enough to be competitive with the premier surveys in other wavelengths (e.g. LSST and the \euclid satellite in the optical and near-infrared), but maximising their potential will require designing surveys and observational techniques informed by real data as well as theoretical models. SuperCLASS aims to provide that data to enable this important future science, allowing radio surveys to fulfil their potential in working alongside optical surveys to mitigate each other's observational systematics. Such robustness to systematics is essential in order to converge on the correct model of the Universe, and to understand how structures within it form on both large (cosmological) and small (galaxy) scales.

\section*{Acknowledgments}

This publication uses data generated via the Zooniverse.org platform, development of which is funded by generous support, including a Global Impact Award from Google, and by a grant from the Alfred P. Sloan Foundation.

IH, MLB, SC and BT acknowledge the support of an ERC Starting Grant (grant no. 280127). TH is supported by a Science and Technology Facilities Council studentship. IH acknowledges support from the European Research Council in the form of a Consolidator Grant with number 681431, and from the Beecroft Trust.

CAH acknowledges financial support from the European Union's Horizon 2020 research and innovation programme under the Marie Sk{\l}odowska-Curie grant agreement No 705332.

CMC thanks the National Science Foundation for support through grants AST-1714528 and AST-1814034, and additionally CMC and SMM thank the University of Texas at Austin College of Natural Sciences for support. In addition, CMC acknowledges support from the Research Corporation for Science Advancement from a 2019 Cottrell Scholar Award sponsored by IF/THEN, an initiative of Lyda Hill Philanthropies. 
SMM also thanks the NSF for a Graduate Student Research Fellowship Program (GRFP) Award number DGE-1610403.

The authors also acknowledge the contribution of Paul Howson to the $K$-band CFHT data reduction.

\ih{This research has made use of the VizieR catalogue access tool, CDS, Strasbourg, France (DOI : 10.26093/cds/vizier). The original description  of the VizieR service was published in 2000, A\&AS 143, 23}

\bibliography{superclass}
\bibliographystyle{mn2e_plus_arxiv}

\label{lastpage}

\end{document}